\documentclass[doublecol]{epl2} 

\usepackage{graphicx,color}

\usepackage{amsmath}
\usepackage[bbgreekl]{mathbbol}
\usepackage{bm}
\usepackage{amssymb}
\usepackage{hyperref}
\usepackage{cite}
\usepackage{xcolor}
\hypersetup{colorlinks,bookmarksopen,bookmarksnumbered,
citecolor=red,
linkcolor=blue,
pdfstartview=green,
urlcolor=teal}

\newcommand{\be}{\begin{equation}}
\newcommand{\ee}{\end{equation}}
\newcommand{\bea}{\begin{eqnarray}}
\newcommand{\eea}{\end{eqnarray}}

\newcommand{\up}{\uparrow}

\title{Relaxation of the order-parameter statistics and dynamical confinement}
\author{Riccardo Javier Valencia Tortora\inst{1}, Pasquale Calabrese\inst{2,3} \and Mario Collura\inst{2} }
\shortauthor{R. J. V. Tortora, P. Calabrese, M. Collura}

\institute{                    
  \inst{1} Institut f\"ur Physik, Johannes Gutenberg Universit\"at Mainz, D-55099 Mainz, Germany\\
  \inst{2} International School for Advanced Studies (SISSA), via Bonomea 265, 34136  Trieste, Italy\\
  \inst{3} International Centre for Theoretical Physics (ICTP) and INFN, I-34151, Trieste, Italy 
}

\abstract{
We study the relaxation of the local ferromagnetic order in the quantum Ising chain in a slant field with both longitudinal and transverse components.
After preparing the system in a fully polarised state, we analyse the time evolution of the entire probability distribution function (PDF) of the magnetisation within a block 
of $\ell$ spins.
We first analyse the effect of confinement on the gaussification of the PDF for large $\ell$, showing that the melting of initial order is suppressed when 
the longitudinal field is aligned to initial magnetisation while it is sped up when it is in the opposite direction. 
Then we study the thermalisation dynamics.  In the paramagnetic region, the PDF quickly shows thermal features.  
Conversely, in the ferromagnetic phase, when confinement takes place, the relaxation suffers a typical slowing down which depends 
on the interplay between the strength of the longitudinal field, the density of excitations, and the direction of the initial polarisation.
Even when the initial magnetisation is aligned oppositely to the longitudinal field, confinement prevents thermalisation in the accessible timescale, 
as it is neatly bared by the PDF. 
}

\begin{document}

\maketitle

\section{Introduction}
Isolated many-body quantum systems
are fundamental theoretical and experimental laboratories to benchmark 
our understanding of elementary processes in nature. 
When brought out-of-equilibrium, they may show 
diametrically opposed behaviour depending on various aspects
that come into play. Low-dimensionality, quantum integrability, and
disorder are just a few  physical features that induce specific responses of the system, 
and therefore affect the relaxation dynamics toward the local steady state
(see the reviews~\cite{bib:pssv11,bib:ge16,bib:vr16,bib:ef16,cem-16}).

Recently, confinement of quantum excitations 
was brought to the fore as one mechanism to 
explain a number of unexpected non-equilibrium phenomena 
in condensed-matter physics~\cite{bib:kctc17,bib:jkr19,bib:rjk19,bib:mplcg19,bib:lzsg19,bib:lltpzmg19,
bib:lsmpcg_arxiv,bib:pdrzm16,bib:mdfppe_arxiv,bib:czlt_arxiv,bib:pp20,bib:bch11,cr-19,exp-20,vlwgh-20,ylgi-20,mrw-17,vk-20,clsv-20}.
For example, %in the contest of spin systems,
when a quantum Ising chain in a transverse field is brought out of equilibrium,  
excitations propagate ballistically and correlations spread accordingly, leading to the renowned {\it light-cone effect}~\cite{cc-06}.
However, with an additional  longitudinal field in the ordered phase, 
the system experiences a completely different behaviour because
excitations are confined into {\it mesons}~\cite{bib:kctc17}.

It is worth to investigate the effects of the confinement  within the
genuinely quantum mechanical framework of the
Probability Distribution Function (PDF), 
which gives a complete characterisation of the 
associated observable, thus going much beyond the mere averages.
PDFs have been investigated in many contexts, but
in spite of the large literature regarding equilibrium setups~\cite{bib:gadp06,bib:lp08,bib:ia13,bib:sk13,bib:e13,bib:er13,bib:lddz15,bib:sp17,bib:ceg17,bib:nr17,bib:hb17,bib:ag19,bib:nr20,bib:bag20,ccgm-20,cd-07,msc-15}, 
results on non-equilibrium states are still scarce~\cite{bib:gec18,bib:bpc18,bib:bp18,bib:ppg19,bib:c19,bib:ce20,bib:cdcd20}.

Actually, by analysing the PDF dynamics,  
we address the following questions altogether:
Is there a remnant local order enhanced by the confinement of excitations?
Does it depend on the explicit breaking of the $\mathbb{Z}_{2}$ symmetry? 
Can confinement affect thermalisation  and ferromagnetic order
in the thermodynamic limit?

The Letter is organised as follows: 
First we introduce the model and briefly review some 
of its features; we then define the observables of interest 
and outline the dynamical protocol. 
After few details on the numerical technique,
we focus on the results of the PDF dynamics. % for confined and non-confined regimes.
Finally we draw our conclusions.

\section{Model}
We consider the quantum Ising spin chain with transverse ($h_z$) and longitudinal ($h_x$) fields, described by 
the Hamiltonian 
\be\label{eq:H}
H = - \sum_{j} \sigma^{x}_{j} \sigma^{x}_{j} 
- h_z \sum_{j} \sigma^{z}_{j}
- h_x \sum_{j} \sigma^{x}_{j},
\ee
where $\sigma^{\alpha}_{j}$ are Pauli matrices acting on the site $j$.
When $h_x = 0$ the model can be mapped to non-interacting spin-less fermions,
and it is characterised by: {\it (i)} a symmetry broken phase (for $|h_z|<1$) with non-zero
order parameter $\langle \sigma^{x}_{j} \rangle$ \cite{sach-book};
{\it (ii)} a paramagnetic phase (for $|h_z|>1$) without magnetic order \cite{sach-book};
{\it (iii)} a non-equilibrium dynamics driven by stable excitations which
ballistically propagates throughout the system \cite{cef-11}.
Conversely, when $h_x\neq 0$, the model is no more integrable, 
and the quasi-particles may experience a different fate depending on the 
presence of confinement~\cite{bib:kctc17,bib:bch11}.
Such effect takes place within the ferromagnetic region of the unperturbed Ising model.
To understand the mechanism, 
let us focus on the quasi-classical regime 
where $|h_z| \ll 1$ and the magnetic fluctuations are small: 
here the excitations are dilute, and correspond to freely moving kinks 
connecting regions with opposite magnetisation $\langle \sigma^{x}_{j}\rangle \neq 0$.
By adding a finite $h_x$, the many-body spectrum %of the Hamiltonian 
is non-perturbatively modified: 
the true ground state is the one polarised along the magnetic field and the energy of a domain in the 
opposite direction increases linearly with its size \cite{mw-78}.
This originates an effective strong interaction between consecutive kinks 
which therefore get confined in a finite spatial region, 
similarly to mesons in quantum chromodynamics.

%In the following we study the effects induced by the longitudinal field 
%in the dynamics of the probability distribution function of the order parameter.

\section{Full counting statistics}
We are interested in the probability distribution function,
or Full Counting Statistics (FCS),
$
P_{\ell}(m) = \langle \delta(M_{\ell} - m) \rangle,
$
of the order-parameter in a subsystem of $\ell$ sites
\be
M_{\ell} = \frac{1}{2}\sum_{j=1}^{\ell}\sigma^{x}_{j}.
\ee
Several past studies in a variety of models considered the PDFs of either observables within one (or few) sites or global ones, but for our aims 
it is fundamental to focus on the FCS within a block. 
Indeed, $M_{\ell}$ is the right observable to understand relaxation to a statistical ensemble since on the one hand it generically 
relaxes (differently from global observables) and, on the other, it can have a thermodynamic behaviour in $\ell$ (as opposed to localised ones).

%Thanks to the properties of $M_{\ell}$, 
The PDF $P(m)$ (with $m$ being integer or half-integer depending on the parity of $\ell$) 
is the Fourier transform of the generating function $F_{\ell}(\lambda)=\langle  {\rm e}^{i \lambda M_{\ell}} \rangle$, i.e.
%\be
%P_{\ell}(m) = 
%\widetilde P_{\ell}(m)
%\sum_{r\in\mathbb{Z}} 
%\delta\left(m - r - [1-(-1)^{\ell}]/4\right),
%\ee
%where
\be
P_{\ell}(m) = 
\int_{-\pi}^{\pi}\frac{d\lambda}{2\pi} \, 
{\rm e}^{-i m \lambda} F_{\ell}(\lambda).
\ee

An important feature of the PDF is the {\it gaussification} for for large $\ell$. 
This is easily understood in terms of the cumulants  $C^{(k)}_{\ell} = (-i)^{k}\partial_{\lambda}^{k}\log F_{\ell}(\lambda)|_{\lambda = 0}$.
When the correlation length $\xi$ is finite, all cumulants are extensive in subsystem size, i.e., $C^{(k)}_{\ell}  = \ell c_{k} +o(\ell)$ as soon as 
$\ell\gg\xi$, see, e.g., \cite{f-19}. 
Consequently, all standardised moments (such as skewness and kurtosis) vanish
%the central limit theorem applies, 
and the PDF is asymptotically gaussian 
\be\label{eq:pdf_gauss}
\widetilde P_{\ell}(\mu \ell) \simeq 
\frac{1}{\sqrt{2\pi \ell c_{2}}}  \exp \left[ - \ell \frac{(\mu - c_{1})^2}{2c_{2}} \right].
\ee
%We conclude that gaussification is expected at any time for large enough $\ell$. 
Any deviation from gaussianity is a signature of anomalous behaviour 
like those expected in the post-quench evolution in a confined phase.

Another important aspect concerns {\it thermalisation}. 
For a generic chaotic model, thermalisation should occur in the long time limit \cite{eth,bib:ge16} 
and therefore the PDF should  approach
the characteristic thermal distribution
$
\mathcal{P}_{\ell}(m) 
= {\rm Tr} [ \delta(M_{\ell} - m) \varrho]
$
where $\varrho = {\rm e}^{-\beta H}/{\rm Tr} ({\rm e}^{-\beta H})$
with the inverse temperature $\beta$ fixed by the energy of the initial state. 
It has been proposed that confinement can prevent thermalisation \cite{bib:jkr19,bib:rjk19,bib:lzsg19,bib:mplcg19}, at least for numerical accessible times.
 
In the following we will first discuss how gaussification depends on the quench parameters and how
confinement makes it more difficult. 
Only after we move to thermalisation and to the effects of confinement. 
The two phenomena have some connections, but they are different: gaussification is expected at any time for large $\ell$, 
thermalisation takes place at large time for arbitrary $\ell$.

%Moreover, despite a possible Gaussian restoration, which is nevertheless expected at any time for $\ell\to\infty$, for a generic non-integrable model.
%However, how fast relaxation takes place does depend on the features of the model.

%namely 
%$
%{\rm Tr} (H \varrho) = 
%\langle \Psi(0) | H |\Psi(0)\rangle
%$.

\section{Quench protocol and numerical technique}
%To study the non-equilibrium dynamics of the order-parameter statistic 
We consider the following quench protocol:
{\it (i)} at time $t=0$, the system is prepared in the initial state $|\Psi_0\rangle$ which is fully polarised,
namely 
${|\!\Uparrow\rangle \equiv |\dots \up\up\up\dots \rangle}$ 
%or  ${|\!\Downarrow\rangle \equiv |\dots \down\down\down \dots \rangle}$,
where 
${\sigma^{x} |\!\up \rangle = |\!\up\rangle}$ 
%and ${\sigma^{x} |\!\down \rangle = -|\!\down\rangle}$;
{\it (ii)} thereafter the system evolves unitarily $|\Psi(t)\rangle = \exp(-i H t) |\Psi_0\rangle$ with the  Hamiltonian (\ref{eq:H}).

At $t=0$ the PDF is $\widetilde P_{\ell}(m) = \delta_{m,\ell/2}$;
%depending on the initial polarisation.
the subsequent time evolution strongly depends on the Hamiltonian parameters in particular on 
%Indeed, $h_z$ always triggers the non-equilibrium dynamics, but 
the sign of $h_x$: % distinguishes the two possibile polarisation:  
While for $h_x>0$, $|\Psi_0\rangle$ is a low-energy state of the post-quench Hamiltonian, 
for $h_x<0$, $|\Psi_0\rangle$ falls in middle of the spectrum.

We numerically investigate the real-time dynamics in a finite system
with $L=80$ lattice sites and open boundary conditions. 
At each instant of time the state is described by a normalised 
Matrix Product State (MPS) 
$
|\Psi(t)\rangle = 
\sum_{\sigma_1,\dots,\sigma_L}
{\bf v}^{\dag} \cdot
\prod_{j=1}^{L} 
\mathbb{A}_{j}^{\sigma_j} (t)
\cdot {\bf v} \,
|\sigma_1,\dots,\sigma_{L}\rangle$,
where ${\bf v}_{\alpha} = \delta_{\alpha,1}$ is the boundary vector, 
and $\mathbb{A}_{j}^{\sigma_j}$
are matrices whose (auxiliary) dimension is at most $\chi_{max} =500$.
The time evolution is obtained by employing the 
Time-Evolving Block-Decimation (TEBD) algorithm~\cite{bib:v04}, with second-order
Suzuki-Trotter decomposition of the evolution operator with time step $dt=0.01$.
Doing so, depending on the quench parameters, 
we can reach a maximum 
time $t_{max}\simeq 10 $,
within reasonable truncation errors ($\sim 10^{-6}$).
The MPS representation is very suitable to evaluate the 
generating function $F_{\ell}(\lambda)$, where
$\lambda$ is moved in $[-\pi,\pi]$ with step $d\lambda = 0.01$,
and the subsystem with $\ell$ sites is in the middle of the chain to minimise boundary effects.
$ P_{\ell}(m)$ is eventually obtained by taking the Fourier transform of 
the numerically interpolated $F_{\ell}(\lambda)$.

\begin{figure}[t!]
 \begin{center}
\includegraphics[width=0.24\textwidth]{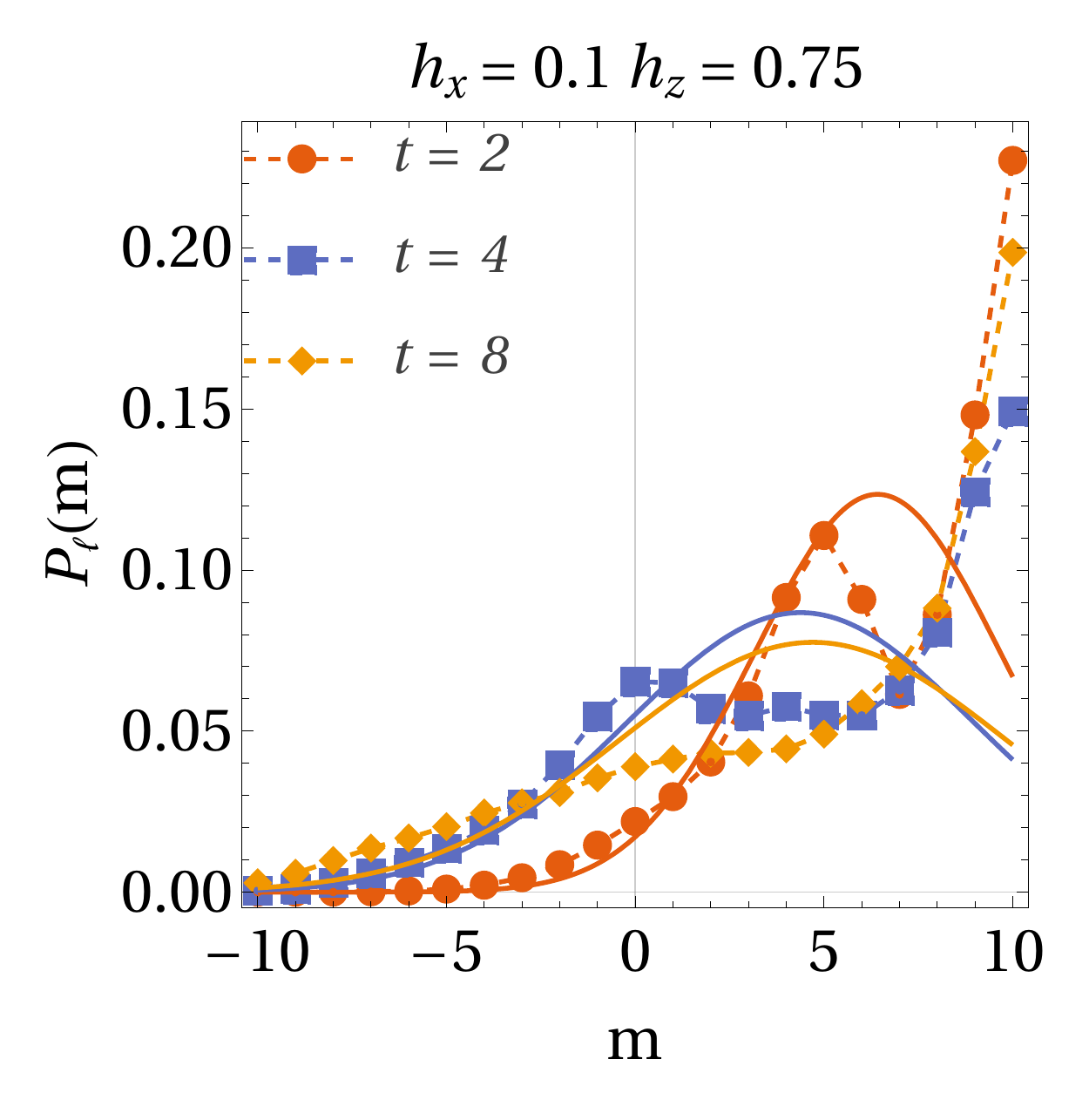} 
\includegraphics[width=0.24\textwidth]{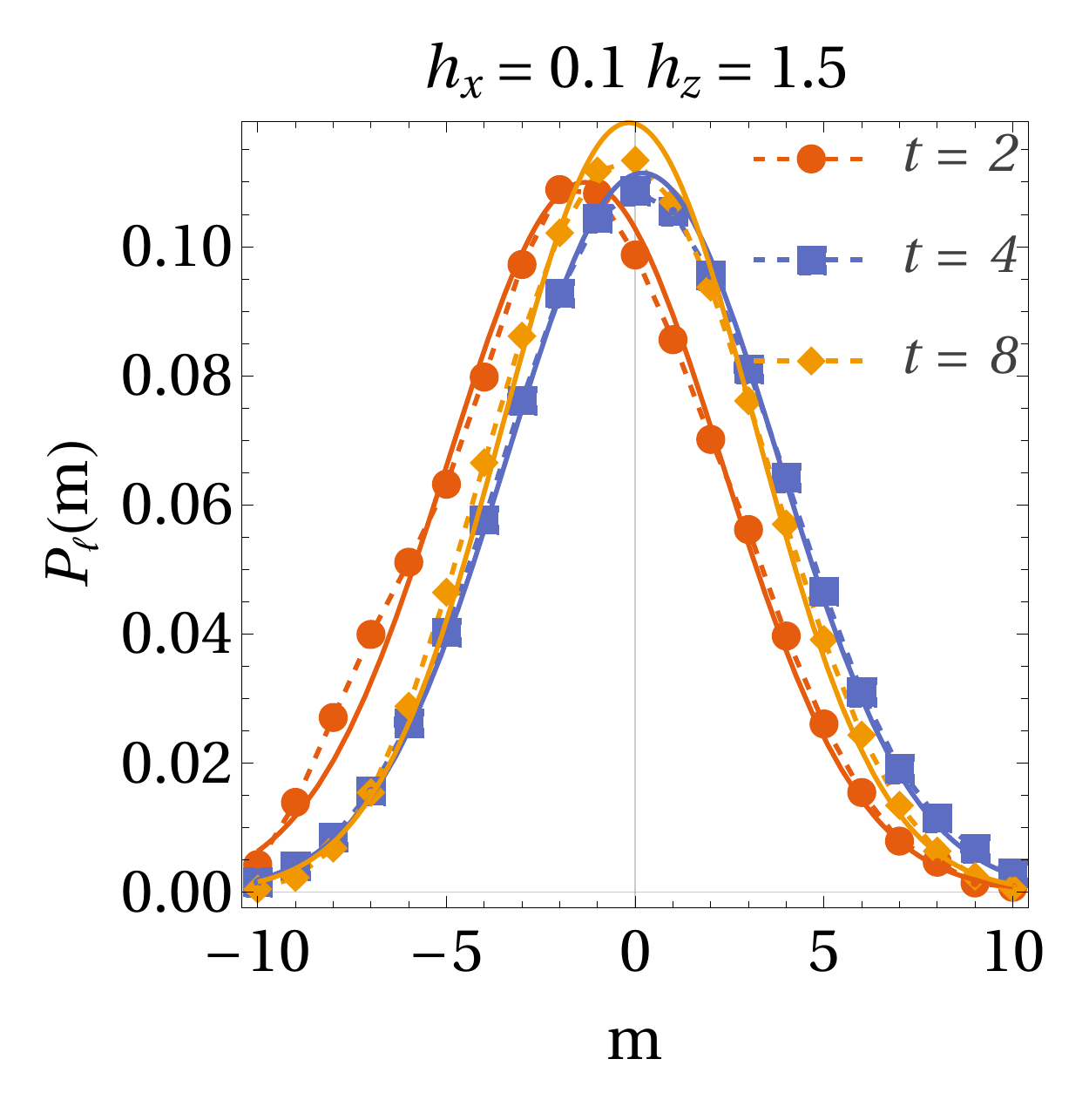} \\
\includegraphics[width=0.24\textwidth]{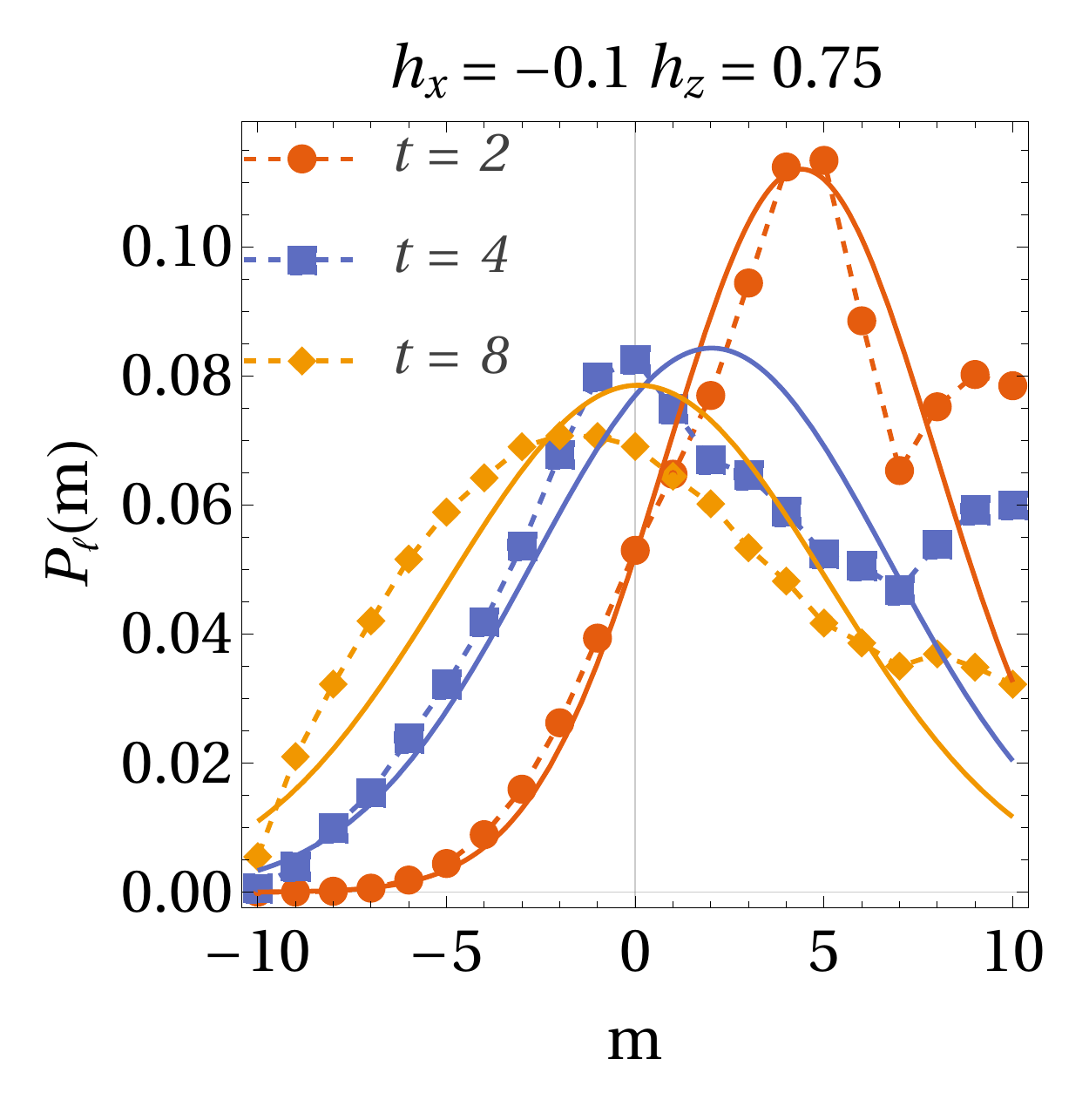} 
\includegraphics[width=0.24\textwidth]{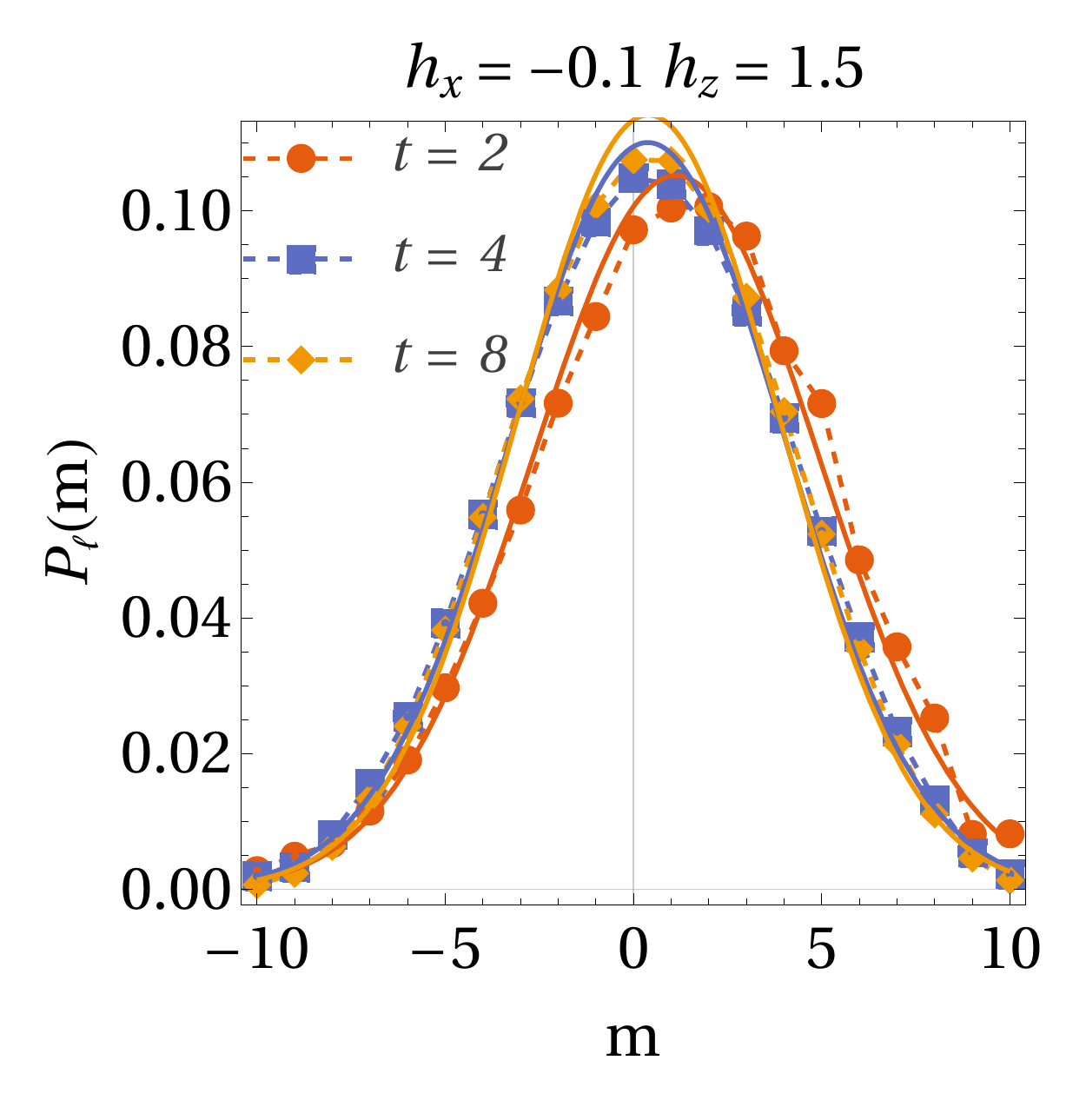} 
\caption{
PDF of the local order parameter with $\ell = 20$ and different times, 
for quenches toward the paramagnetic (right) and ferromagnetic (left) 
phase with $h_x>0$ (top row) and $h_x<0$ (bottom).
TEBD exact results (symbol) are compared with the Gaussian
approximation in Eq.~(\ref{eq:pdf_gauss}) (full lines).
}
\label{fig:pdf_gauss_time}
\end{center}
\end{figure}

\section{Gaussification and memory of the initial order}
Since the quench pumps an extensive energy  into the system,
a finite correlation length should build up during the time evolution (a 1D system cannot have critical points at finite energy density). 
How fast and how large depend on %both the initial state and 
the post-quench parameters $h_z$ and $h_x$.
In general, when the subsystem size is sufficiently larger than
the time-dependent correlation length,
the PDF should fairly match the Gaussian approximation \eqref{eq:pdf_gauss},
where the only parameters are the first two cumulants.
In Fig.~\ref{fig:pdf_gauss_time} we show 
the results for some representative quenches:
For quenches to the paramagnetic phase, 
the time-dependent correlation length is very small and,
for subsystem with $\ell = 20$, we found a
good agreement with the Gaussian~(\ref{eq:pdf_gauss}) at any time;
moreover we do not find any noticeable qualitative difference depending on the sign of $h_x$
or its absolute value.
Indeed, here tuning the Hamiltonian parameters has the only effect 
of changing the characteristic relaxation time.

The PDF dynamics after quenching to the ferromagnetic phase is completely different.
We first discuss what happens for small $|h_x|$ showing in Fig.~\ref{fig:pdf_gauss_time} 
the data for $h_x=\pm0.1$ and up to time $t=8$.
For finite but large $\ell$ (20 in Fig.~\ref{fig:pdf_gauss_time}) the PDF 
is still not Gaussian and shows a very mild dependence on the sign of $h_x$.
For $h_x=0$, in Refs. \cite{bib:c19,bib:ce20}, it has been argued that the deviations from a gaussian are related to
the memory of the initial local order. % whose dynamical slowing-down of the relaxation.
Our conclusion is that, at least for relative short times, the presence of a small longitudinal field 
does not qualitatively alter this effect. %, as clear results for the quenches with $h_x = 0.1$ and $h_z = 0.75$ (see left of Fig.~\ref{fig:pdf_gauss_time}).

%Although a Gaussian PDF is a {\it thermodynamic} feature of a subsystem, 
%here the condition $\ell > \xi$ is by no means sufficient to guarantee a Gaussian behaviour.

\begin{figure}[t!]
 \begin{center}
\includegraphics[width=0.24\textwidth]{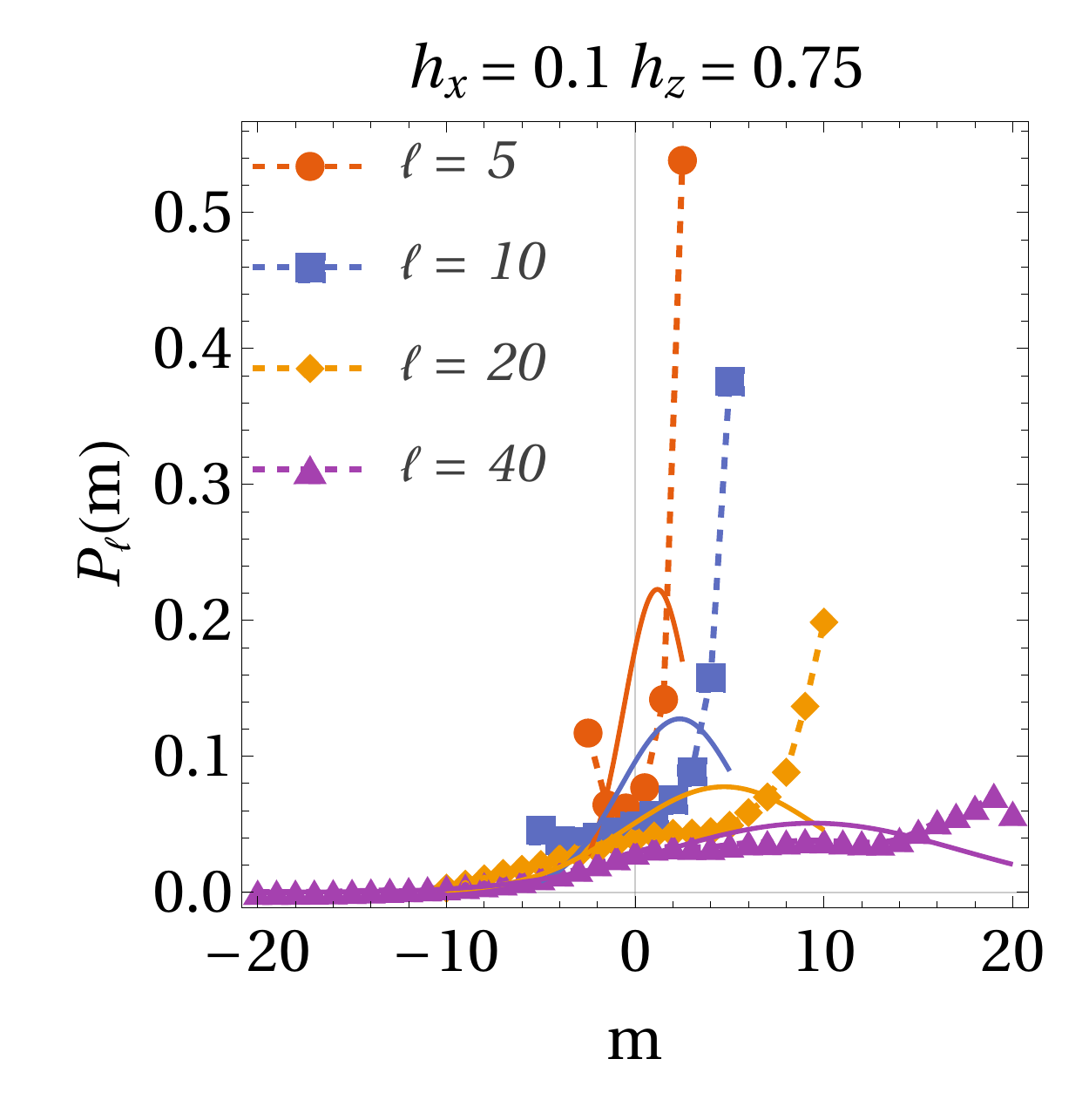} 
\includegraphics[width=0.24\textwidth]{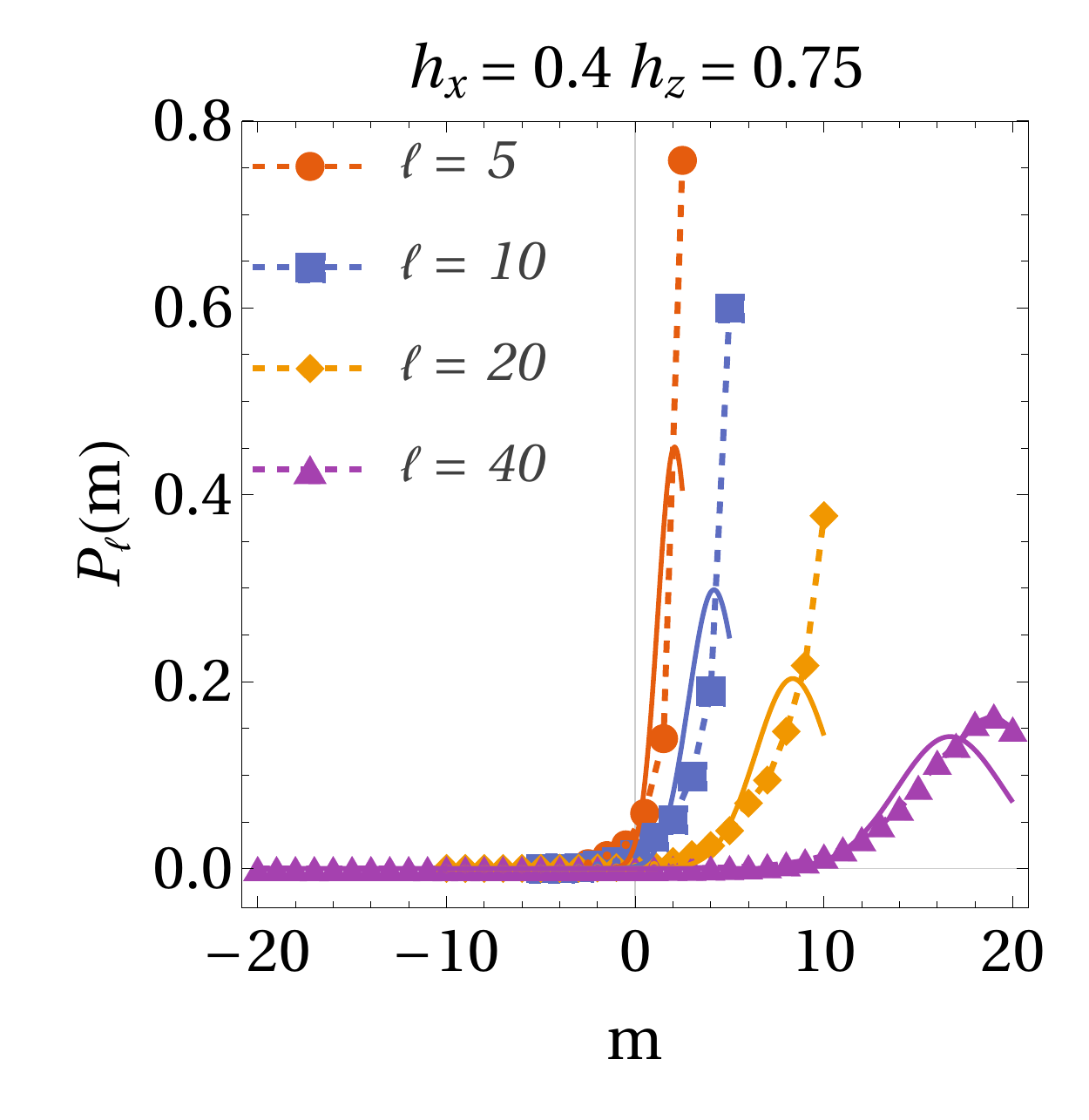}\\ 
\includegraphics[width=0.24\textwidth]{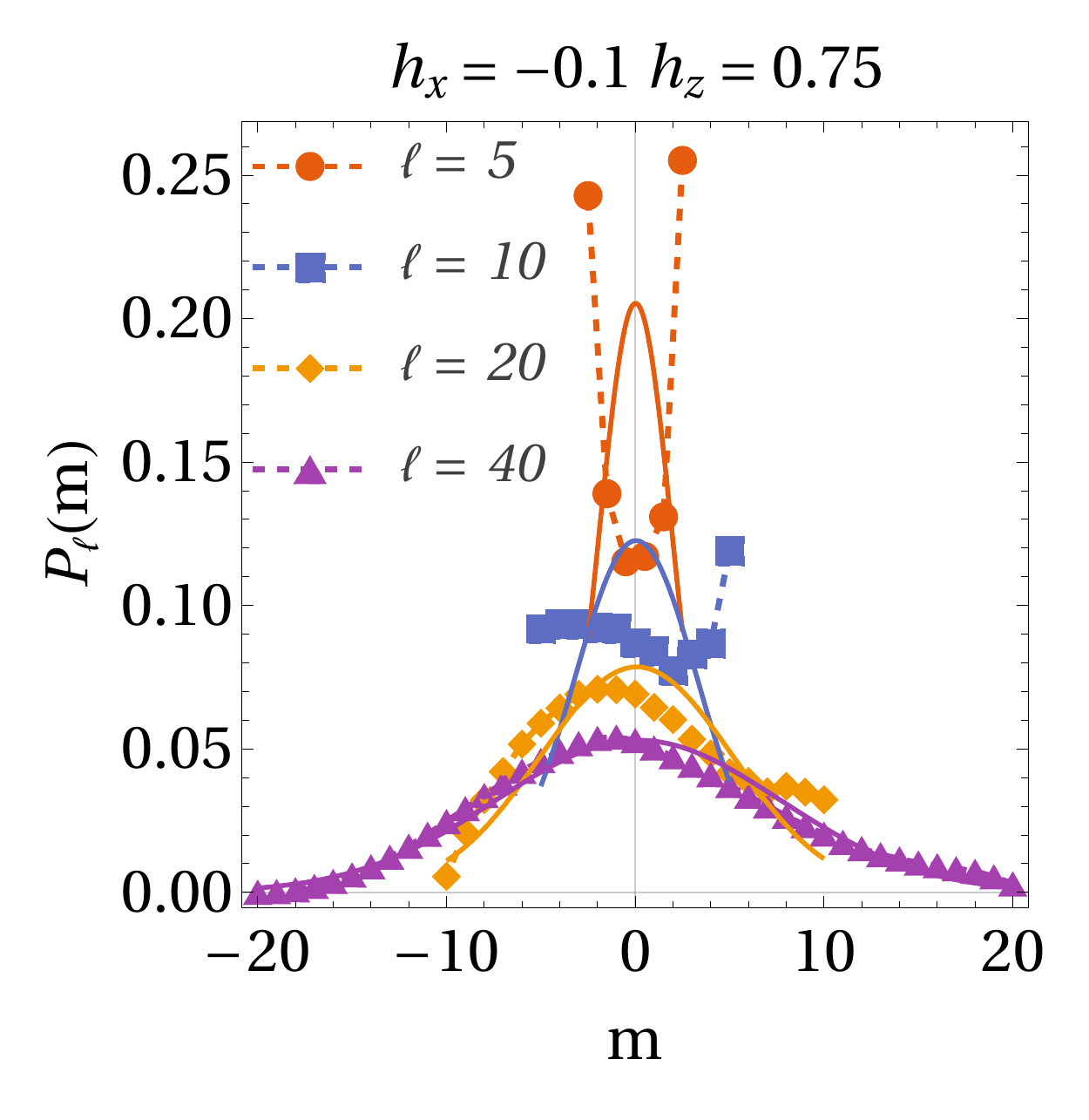} 
\includegraphics[width=0.24\textwidth]{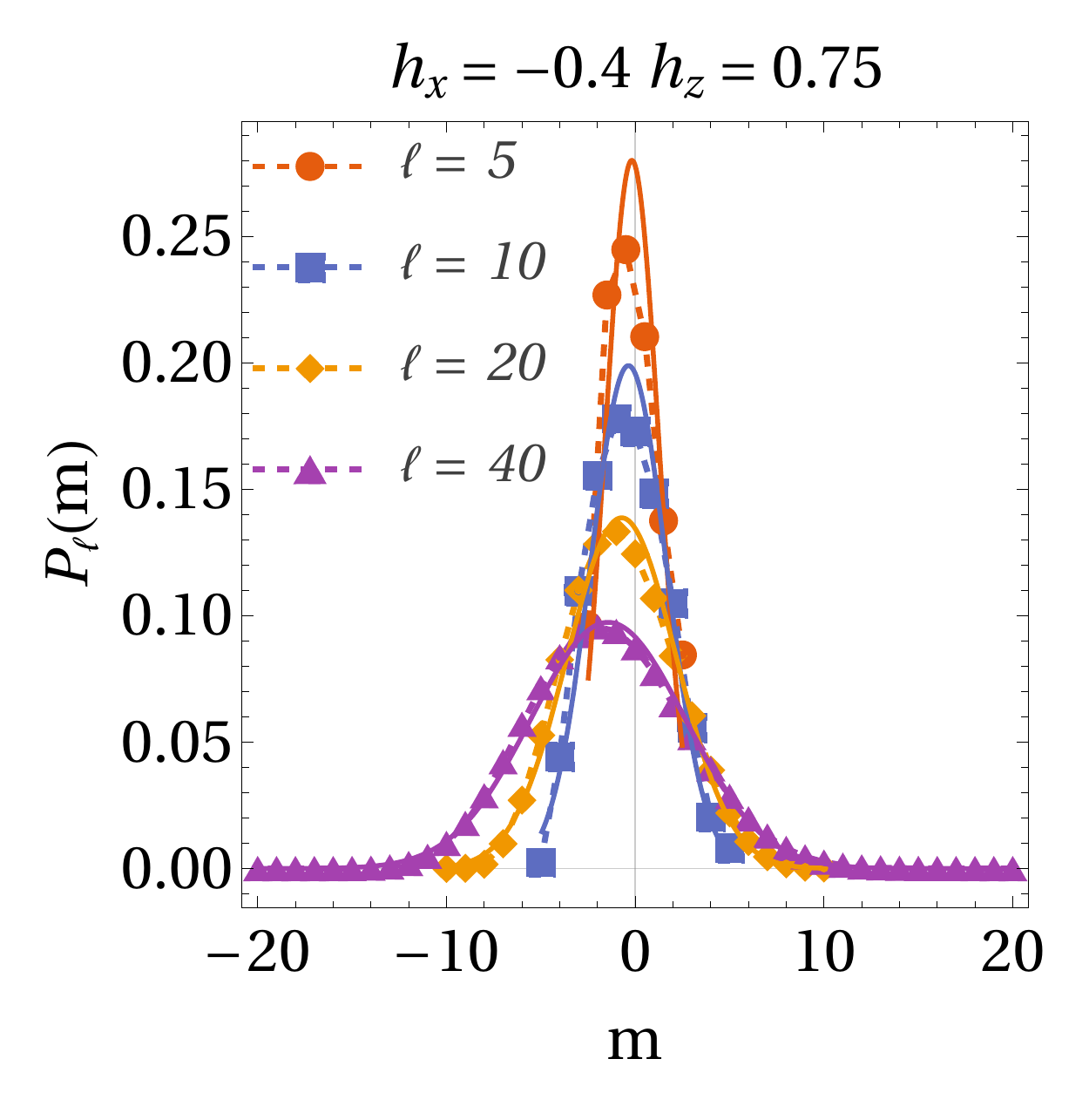} 
\caption{
Order parameter PDF at time $t=8$ for different subsystem sizes $\ell$,
and quenches toward the ferromagnetic phase.
TEBD exact results (symbol) are compared with the Gaussian
approximation in Eq.~(\ref{eq:pdf_gauss}) (full lines).
}
\label{fig:pdf_gauss_ferro}
\end{center}
\end{figure}

\begin{figure*}[t!]
\begin{center}
\includegraphics[width=0.235\textwidth]{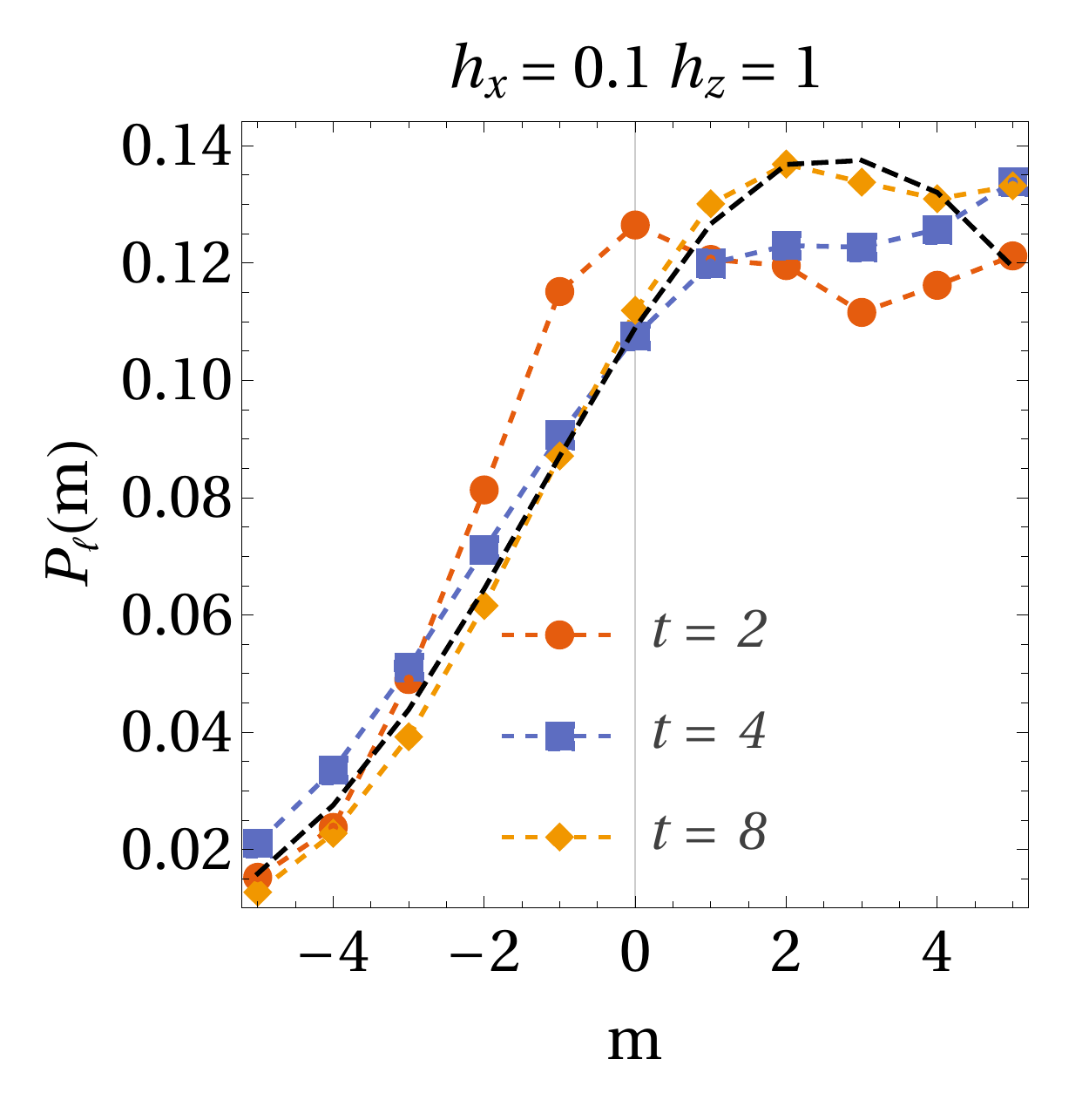} 
\includegraphics[width=0.235\textwidth]{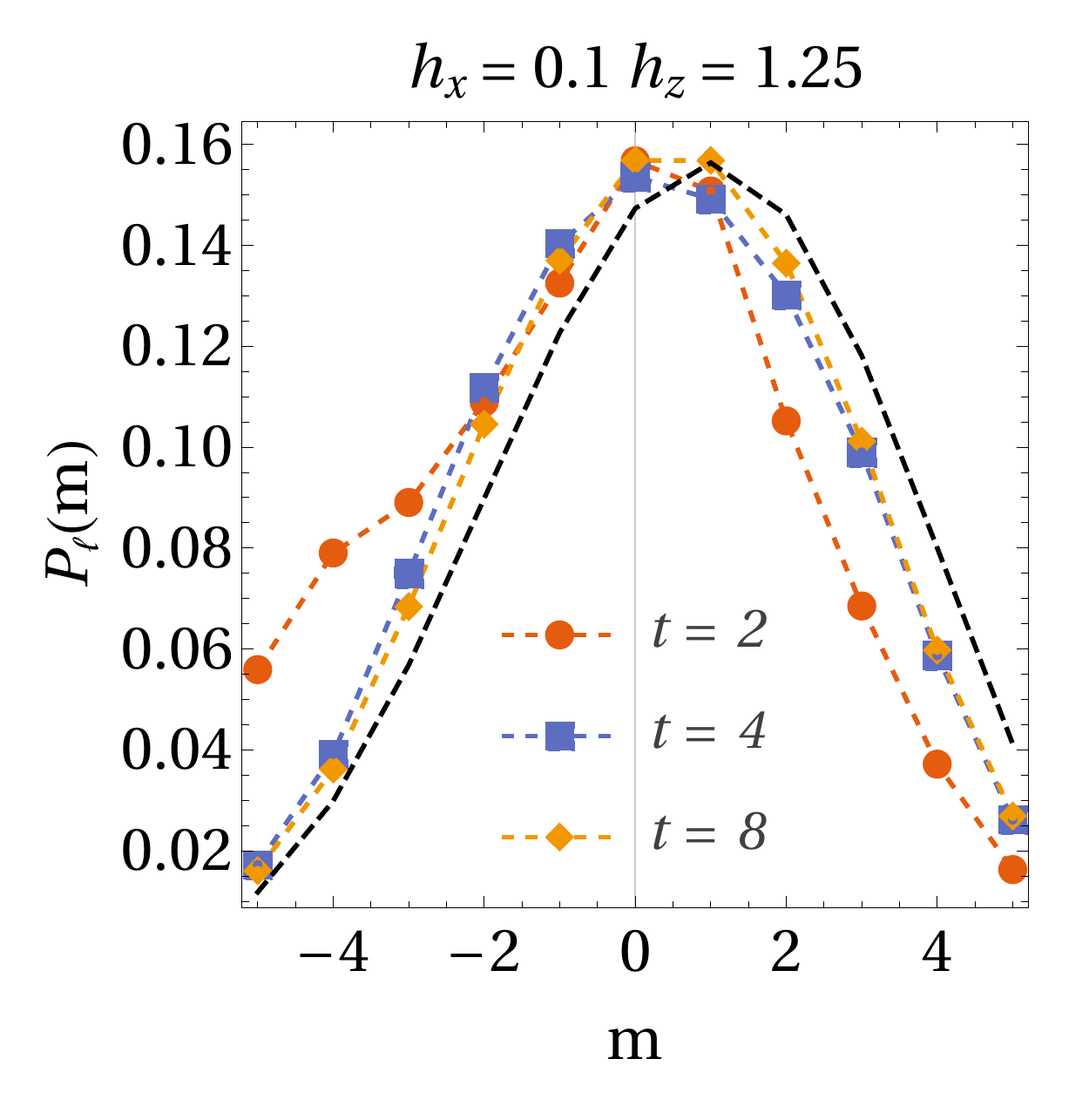}
\includegraphics[width=0.235\textwidth]{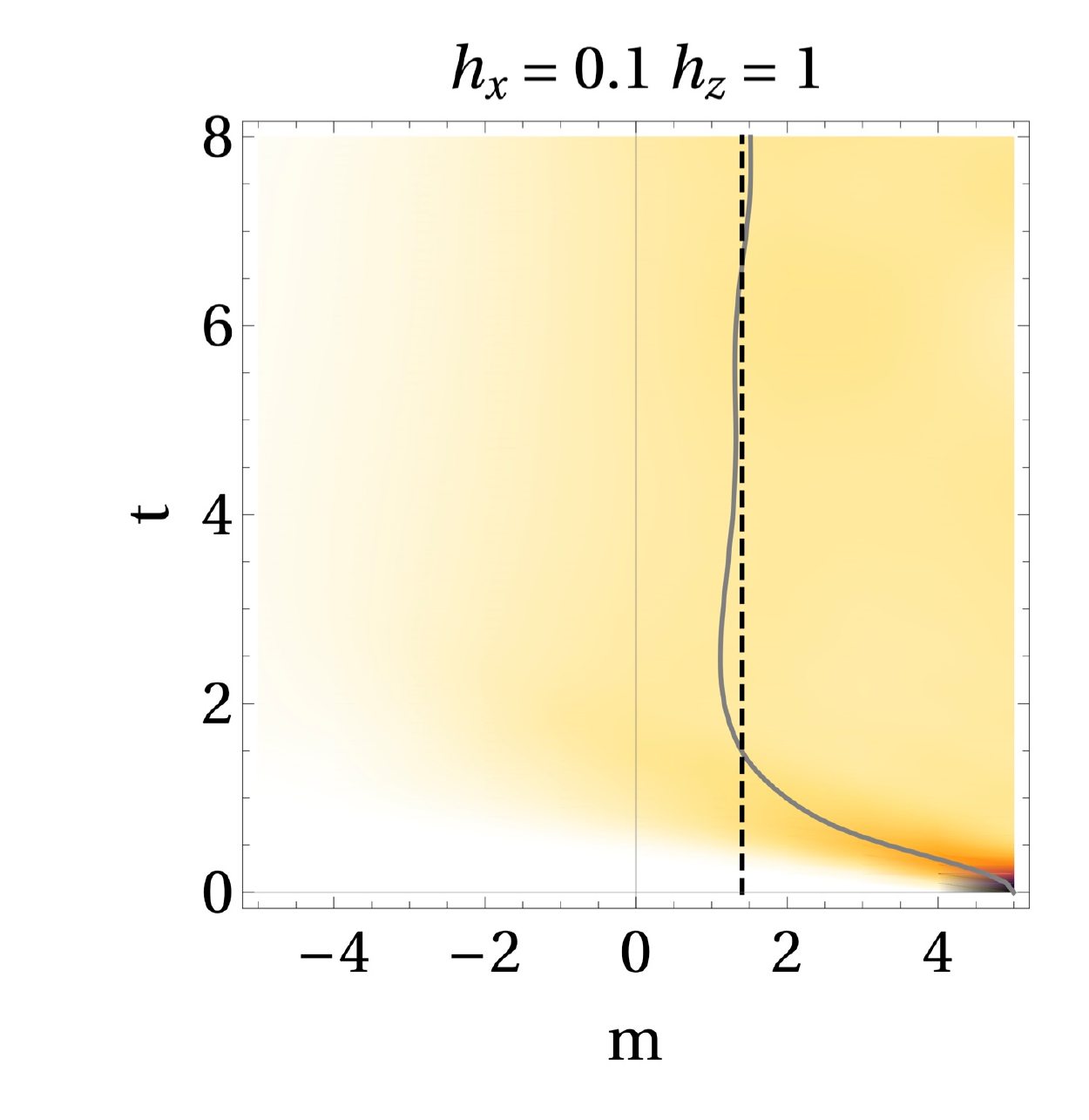}
\includegraphics[width=0.235\textwidth]{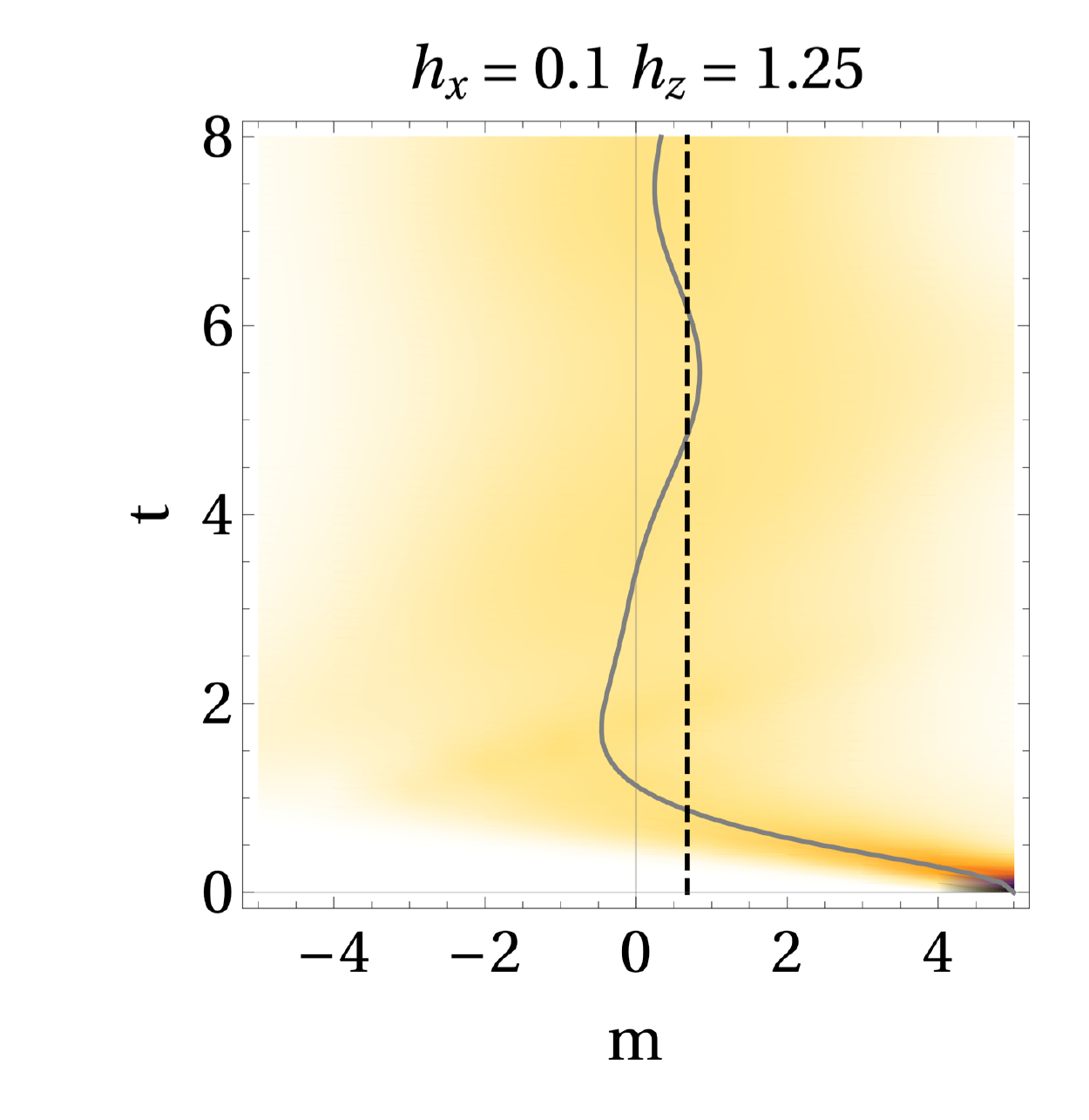}
\includegraphics[width=0.027\textwidth]{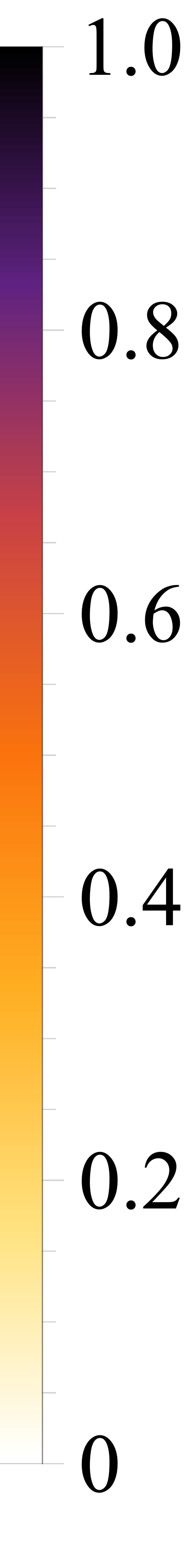}\\
\includegraphics[width=0.235\textwidth]{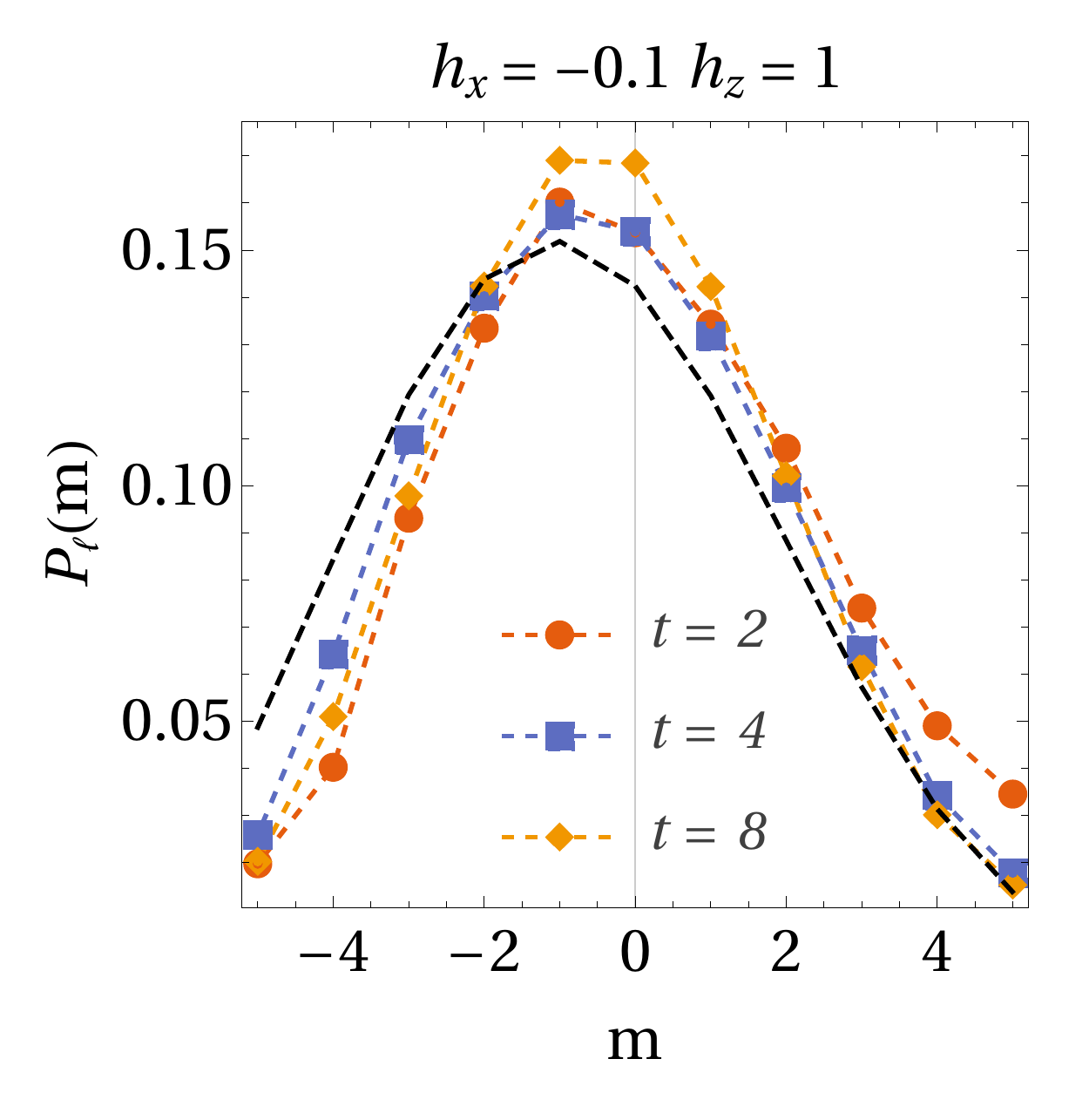} 
\includegraphics[width=0.235\textwidth]{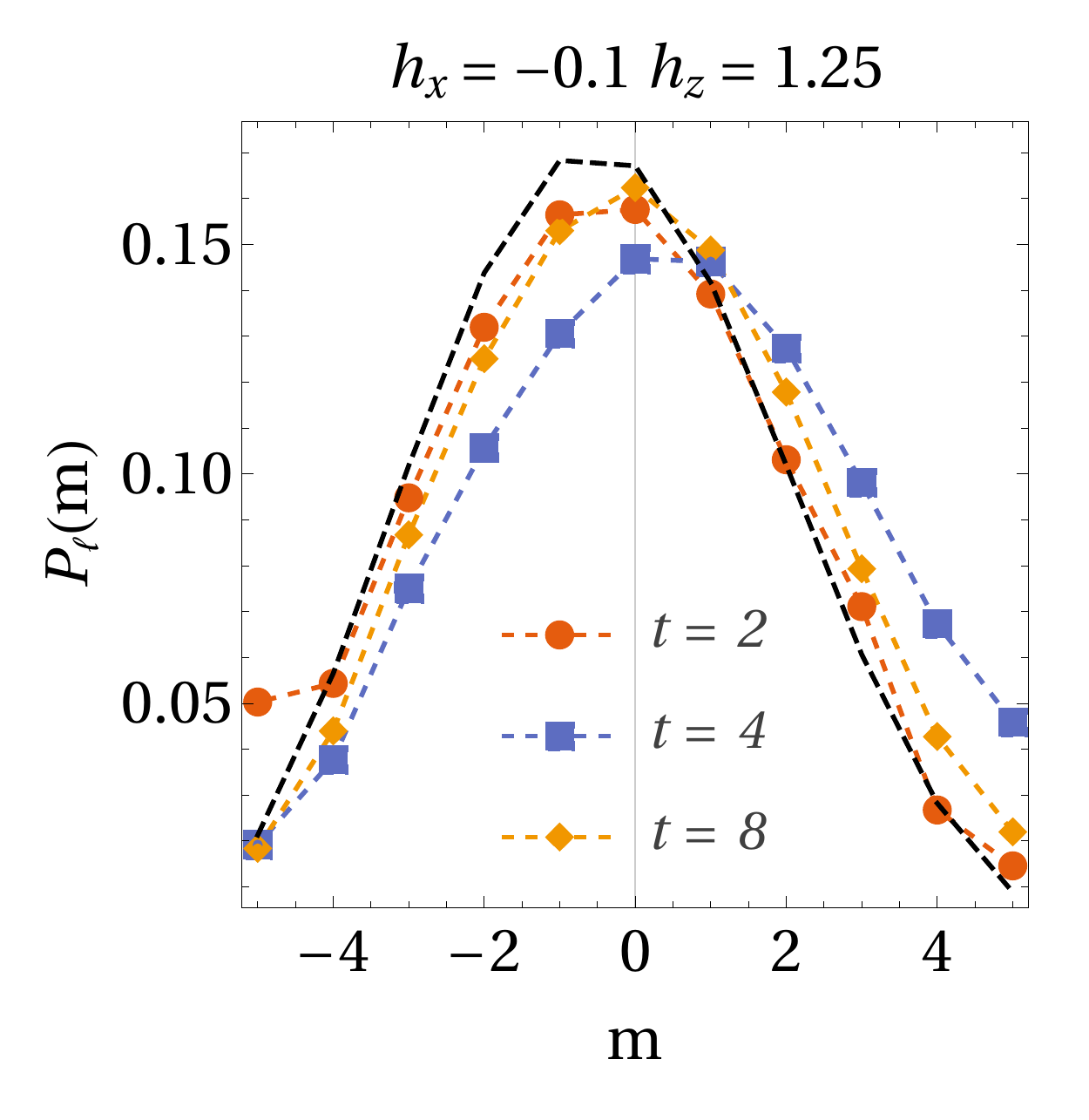}
\includegraphics[width=0.235\textwidth]{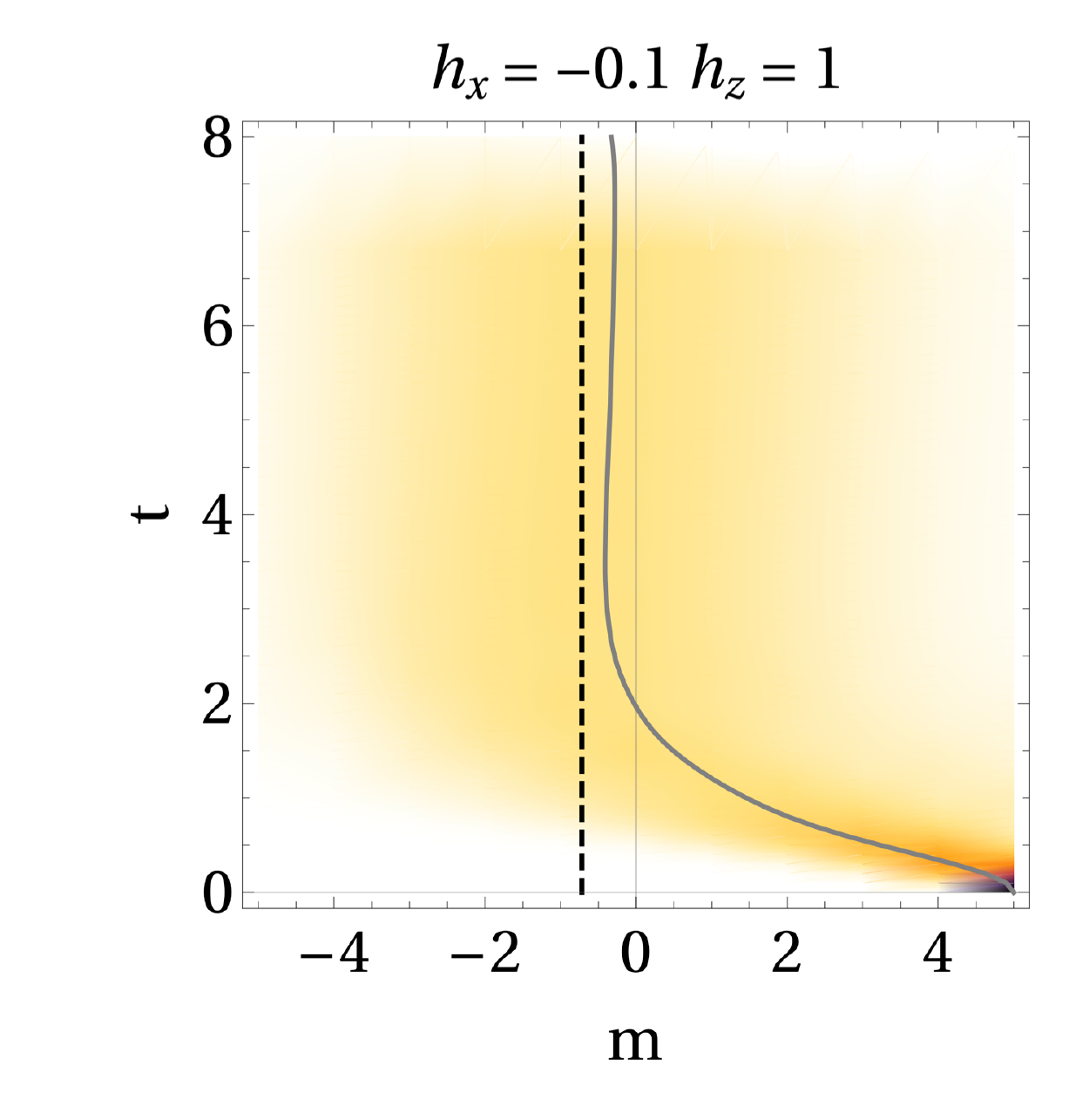}
\includegraphics[width=0.235\textwidth]{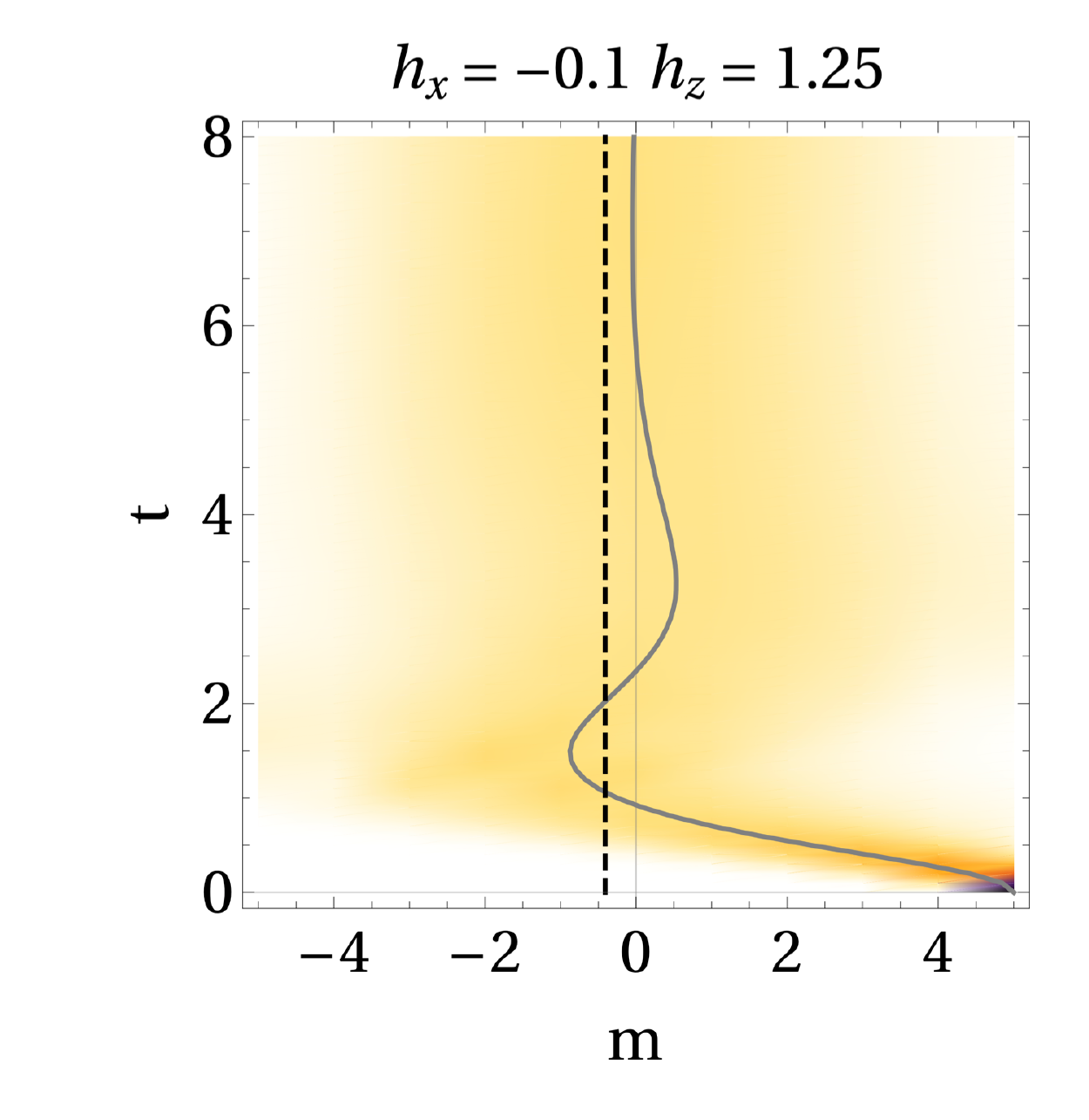}
\includegraphics[width=0.027\textwidth]{figs/fig3/bar.pdf}
\caption{
PDF of the local order parameter for $\ell = 10$ as a function of time for $h_x>0$ (top) and $h_x<0$ (bottom),
for quenches to the paramagnetic and critical phase.
In the two columns on the left, the probability $P_{\ell} (m)$ is drawn as function of $m$ for different times
(symbols) and compared with the stationary thermal distribution (dashed black lines).
In the two columns on the right, we plot the density of the PDF as function of time $t \in [0,8]$ with $m \in[-\ell/2,\ell/2]$. 
The grey full line is the time-dependent magnetisation, whilst the dashed black line is the thermal average.
}
\label{fig:pdf_para}
\end{center}
\end{figure*}

As $|h_x|$ increases, the values of $\ell\gg1$ for which there is gaussification depends strongly on the sign of $h_x$.
In Fig.~\ref{fig:pdf_gauss_ferro} we analyse the PDF for $|h_x|=0.1,0.4$ at fixed time $t=8$ with varying the subsystem size $\ell$.
For $h_x <0$ (i.e. initial polarisation discord to the magnetic field), the PDF is approximately gaussian
for relatively small $\ell$ for $h_x=-0.4$. 
This is not the case for $h_x>0$ when (for $h_x=0.4$) we still observe large deviations from Gaussian at $\ell=40$. 
In other words, the memory of the initial order is enhanced by a strong longitudinal field
in the direction of the initial polarisation; 
instead the local order is easily melted when the longitudinal field is in the opposite direction. 
Microscopically, this behaviour follows from the fact that, by increasing the absolute value of $h_x$, 
while for $h_x<0$ the initial state has larger (and almost uniform) overlaps with eigenstates of the post-quench Hamiltonian
in the middle of the spectrum, 
for $h_x>0$ the larger contributions are from the low-energy states, with small dependence on $h_x$.
%
%Moreover, the value of the longitudinal field allowing Gaussian restoration depends also on the subsystem size.
%This feature is a consequence of the fact that the Hamiltonian (\ref{eq:H}) with $h_x \neq 0$ explicitly breaks the original $\mathbb{Z}_{2}$ symmetry.\pc{??}

\section{Paramagnetic thermalisation and ferromagnetic confinement}
%So far we have only studied how the subsystem PDF acquires a Gaussian character,
%and how this phenomenon depends on the competition 
%between the initial polarisation and the post-quench longitudinal field.
%However, the Hamiltonian (\ref{eq:H}) for $h_x \neq 0$ is chaotic and is expected to thermalise.
%\pc{??Despite Gaussian restoration, {\it thermalisation} is a phenomenon which
%does take place at any length-scale: in other words, for any subsystem size $\ell$,
%provided that $t \gg \ell$, the subsystem PDF should reveal a thermal character,
%which in general for finite $\ell$ is not Gaussian. 
%Moreover, the typical relaxation time does scale differently depending on the properties of the model.}

When quenching to the paramagnetic (and critical) phase, it has been firmly established that the system thermalises in a standard way \cite{bib:bch11}, 
independently of the sign of $h_x$.  
The analysis of the PDF in Fig.~\ref{fig:pdf_para} (for quenches to $h_z = 1$ and $1.25$ with $h_x = \pm 0.1$) confirms this scenario.  
The PDF clearly does what we would expect: The average moves toward the thermal average and the
distribution broadens approaching its thermal expectation in a uniform manner. 
The main reason why we report these expected results here is to contrast them with what happens in the ferromagnetic phase. 
%thermal distribution, namely the PDF calculated in the thermal state $\varrho = {\rm e}^{-\beta H} /Z$ where $\beta$ is fixed by imposing the energy conservation.

\begin{figure*}[t!]
 \begin{center}
\includegraphics[width=0.235\textwidth]{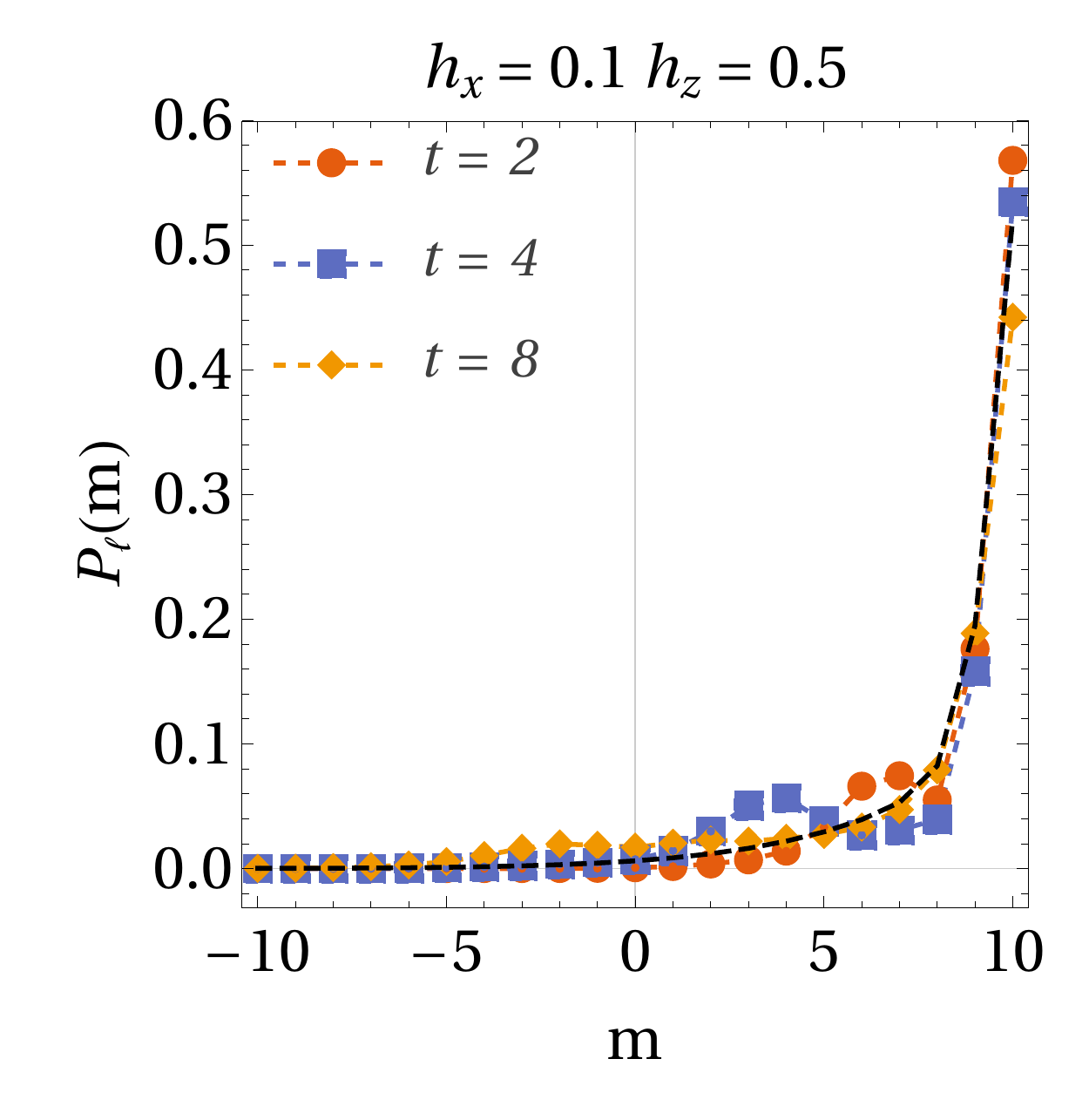} 
\includegraphics[width=0.235\textwidth]{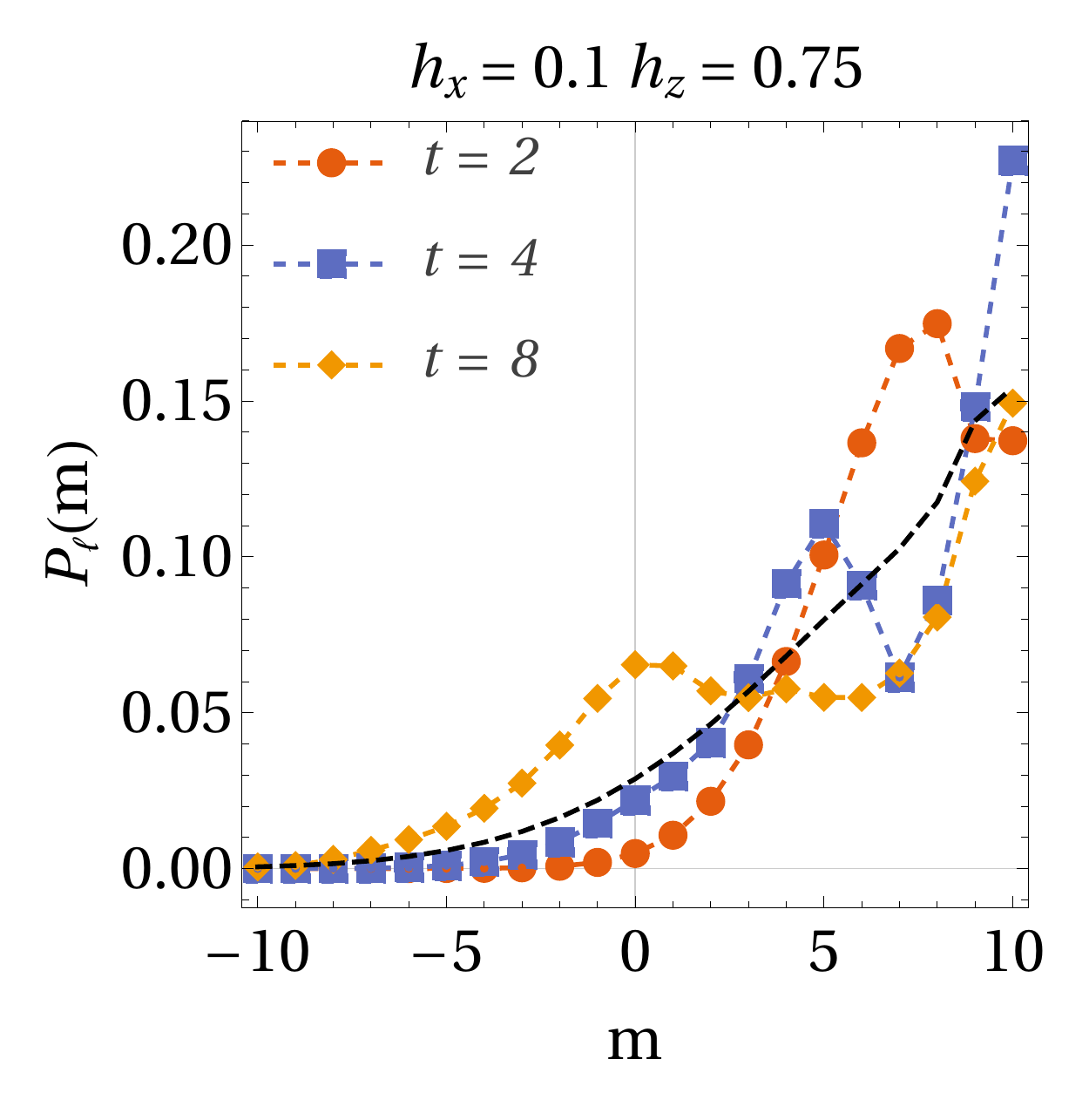}
\includegraphics[width=0.235\textwidth]{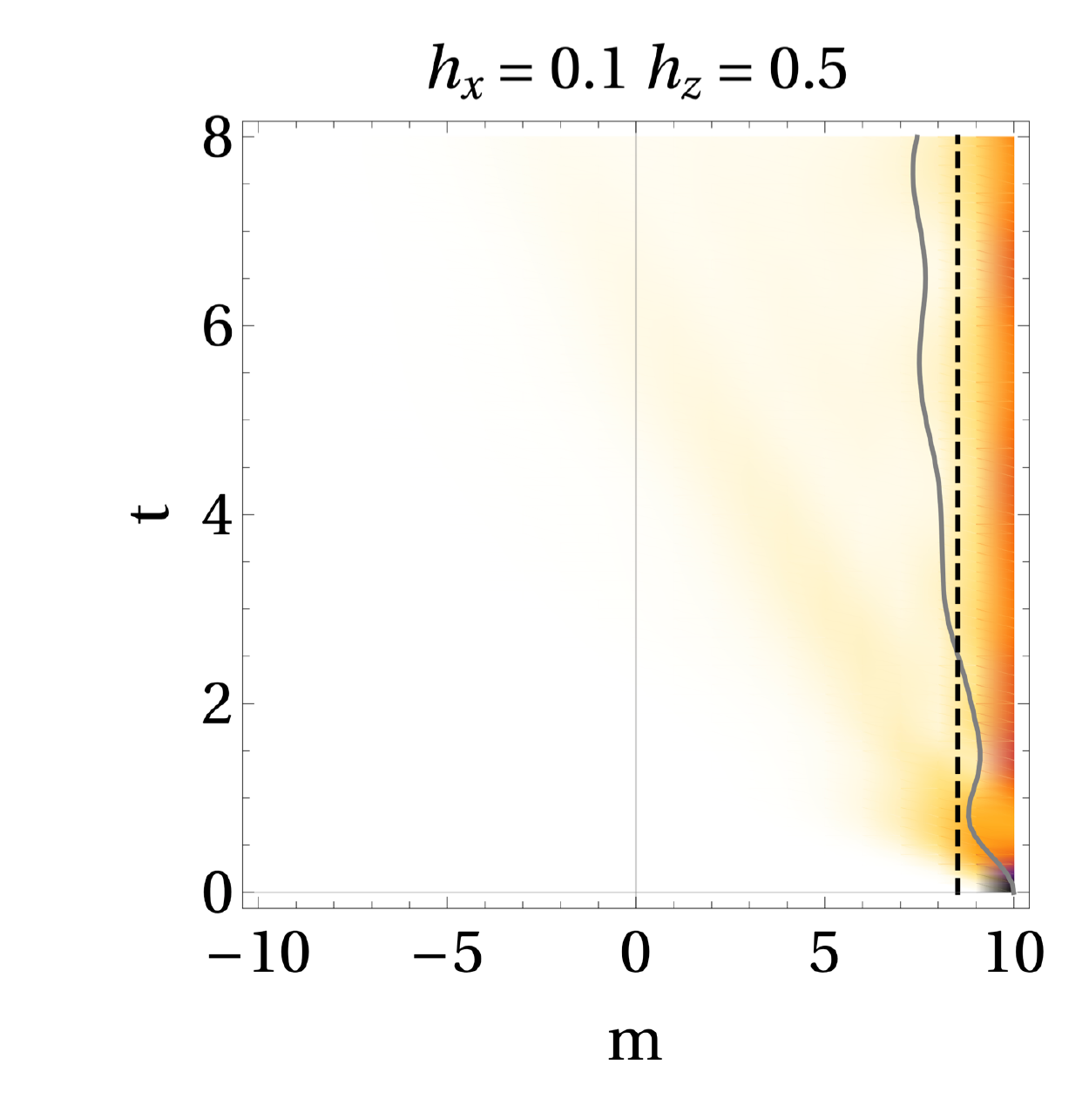}
\includegraphics[width=0.235\textwidth]{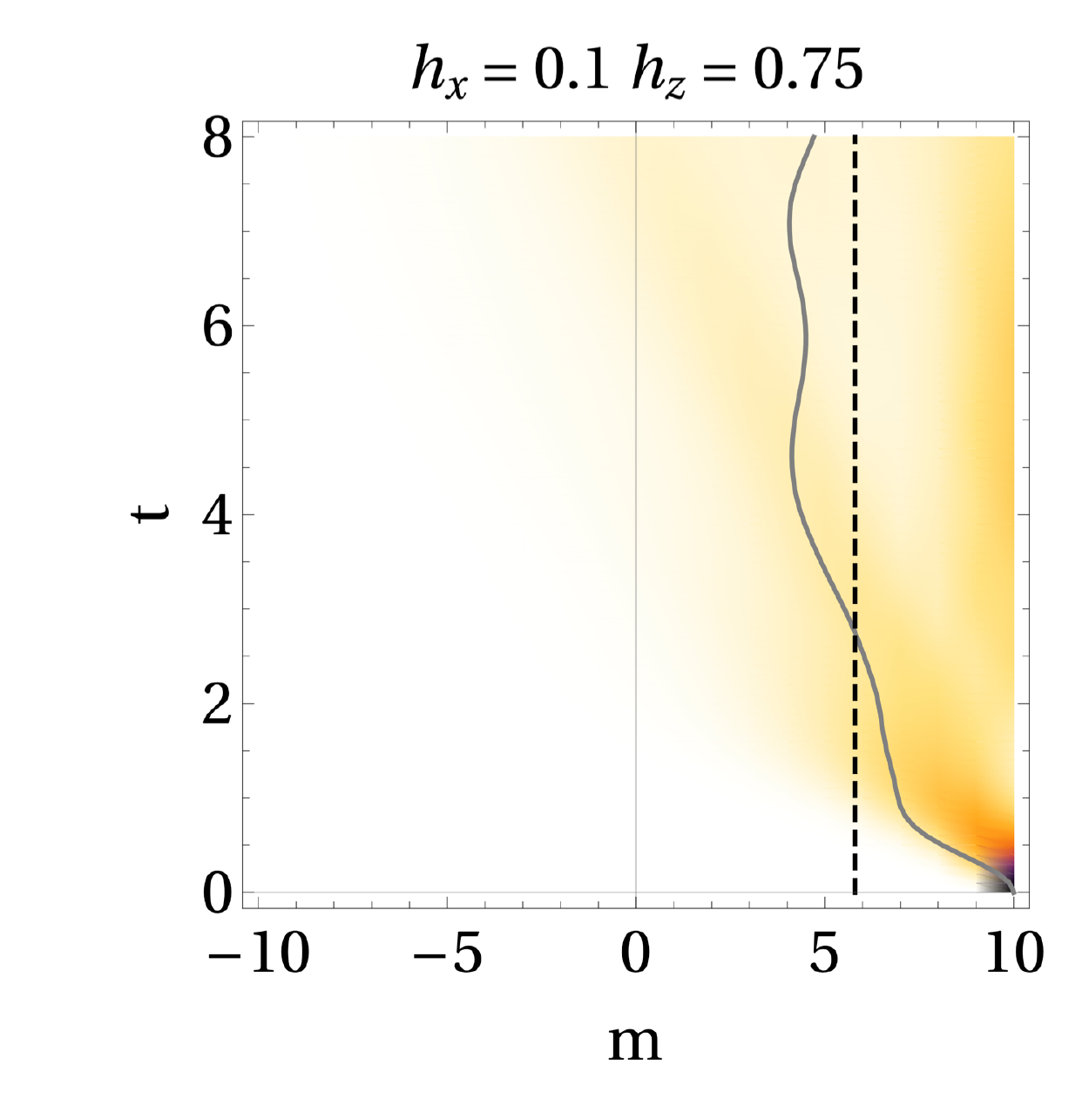}
\includegraphics[width=0.027\textwidth]{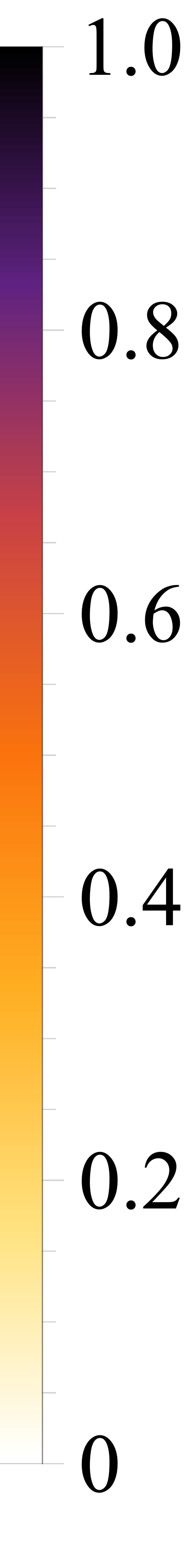}\\
\includegraphics[width=0.235\textwidth]{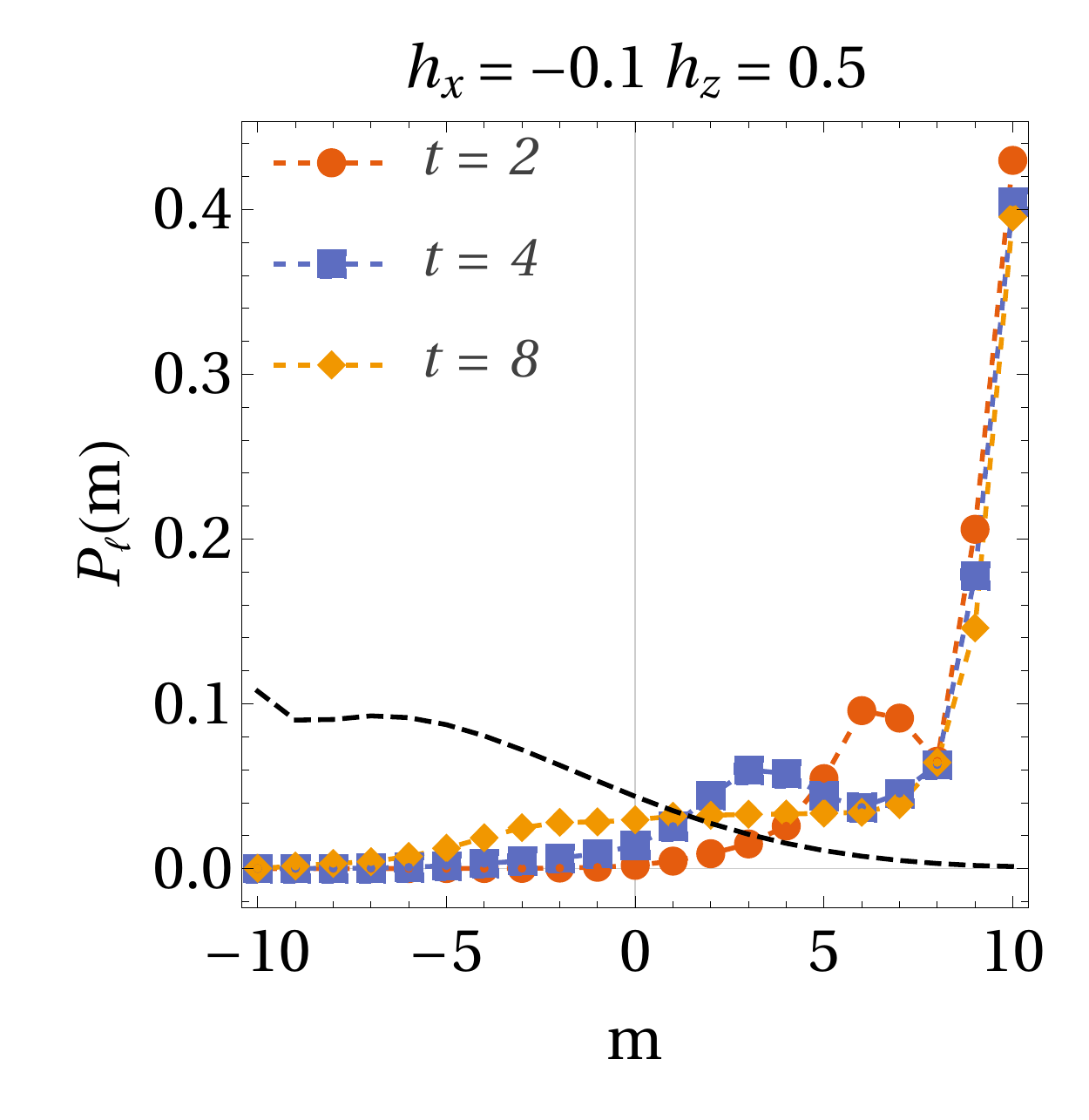} 
\includegraphics[width=0.235\textwidth]{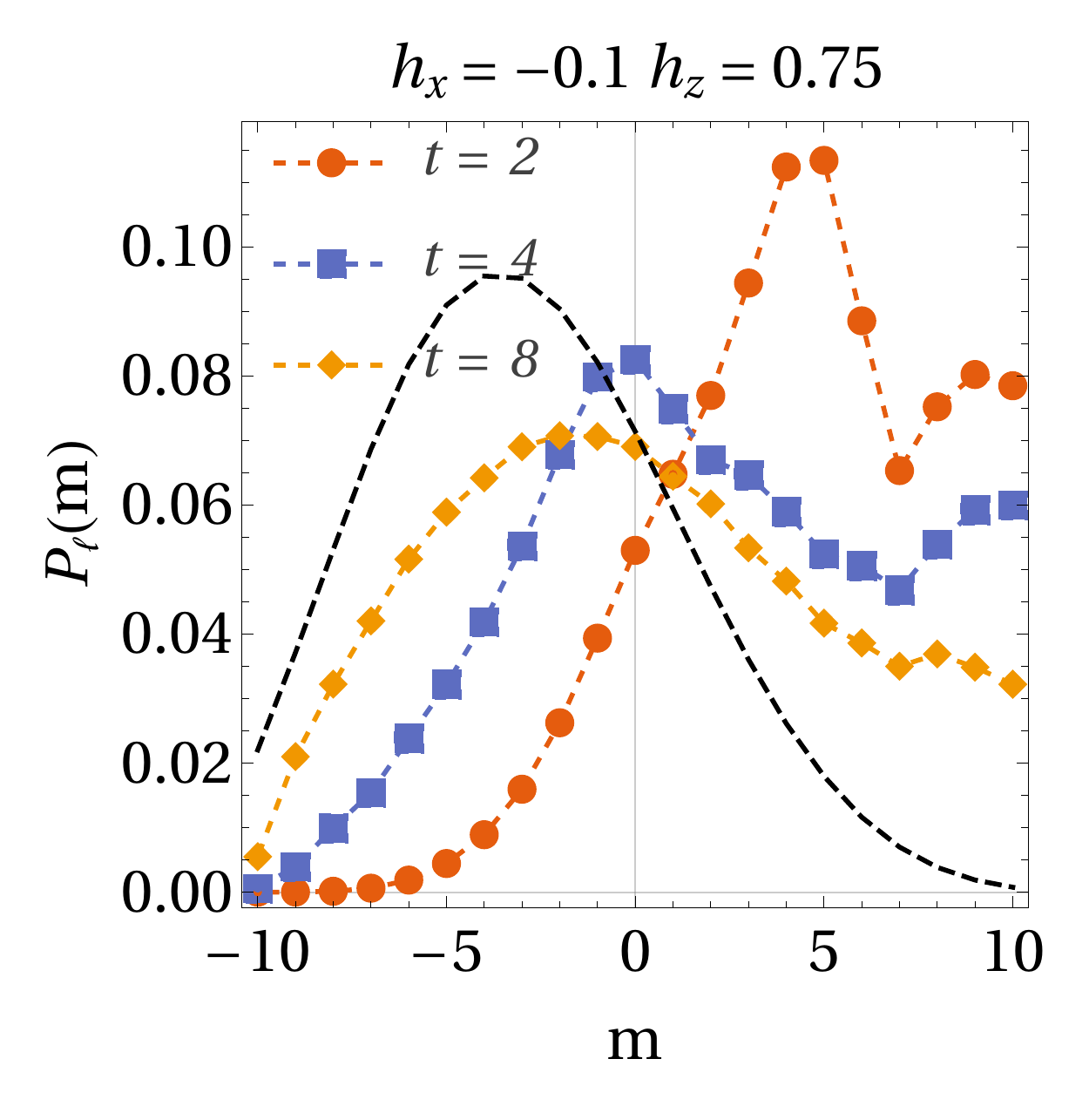}
\includegraphics[width=0.235\textwidth]{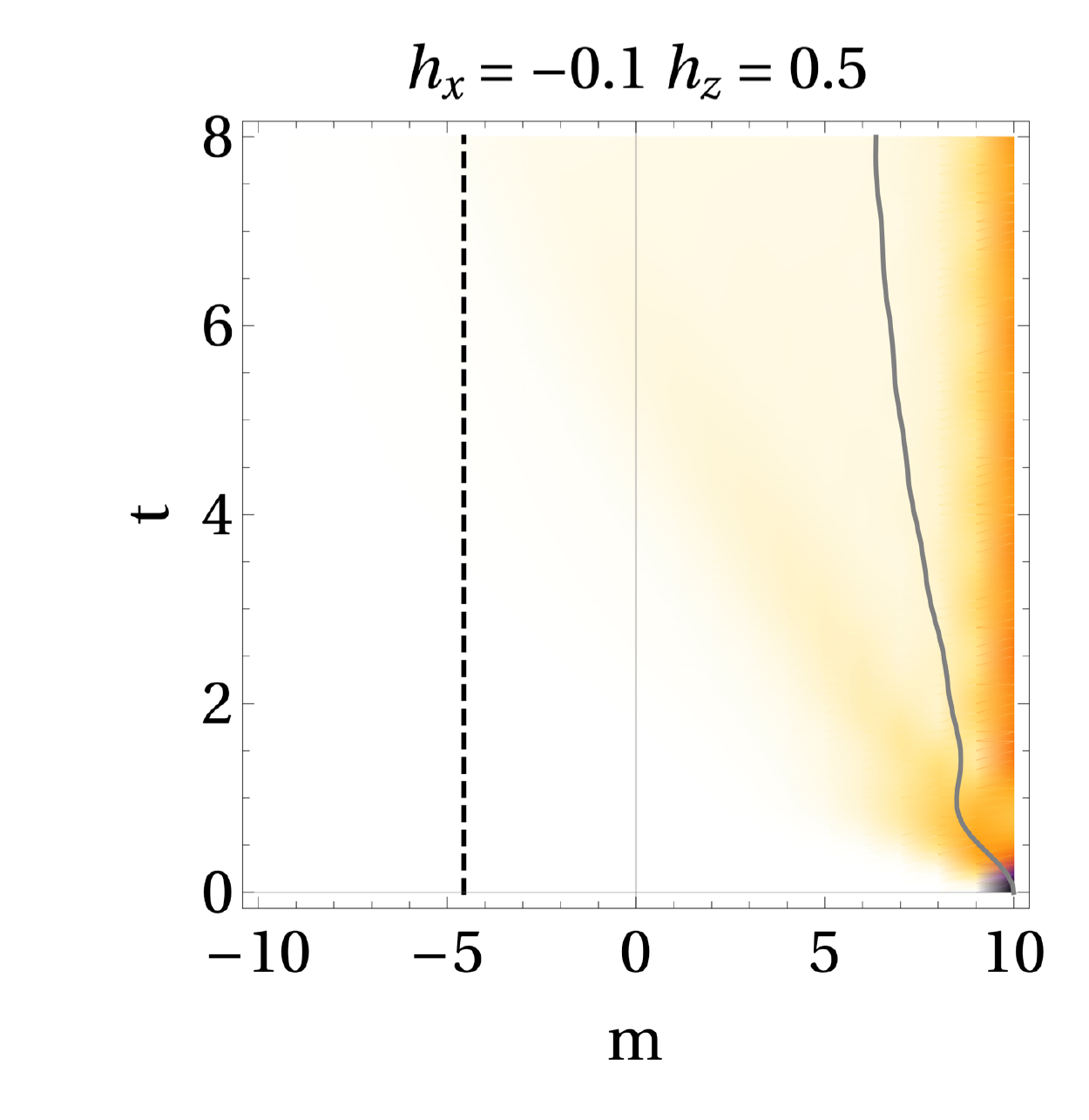}
\includegraphics[width=0.235\textwidth]{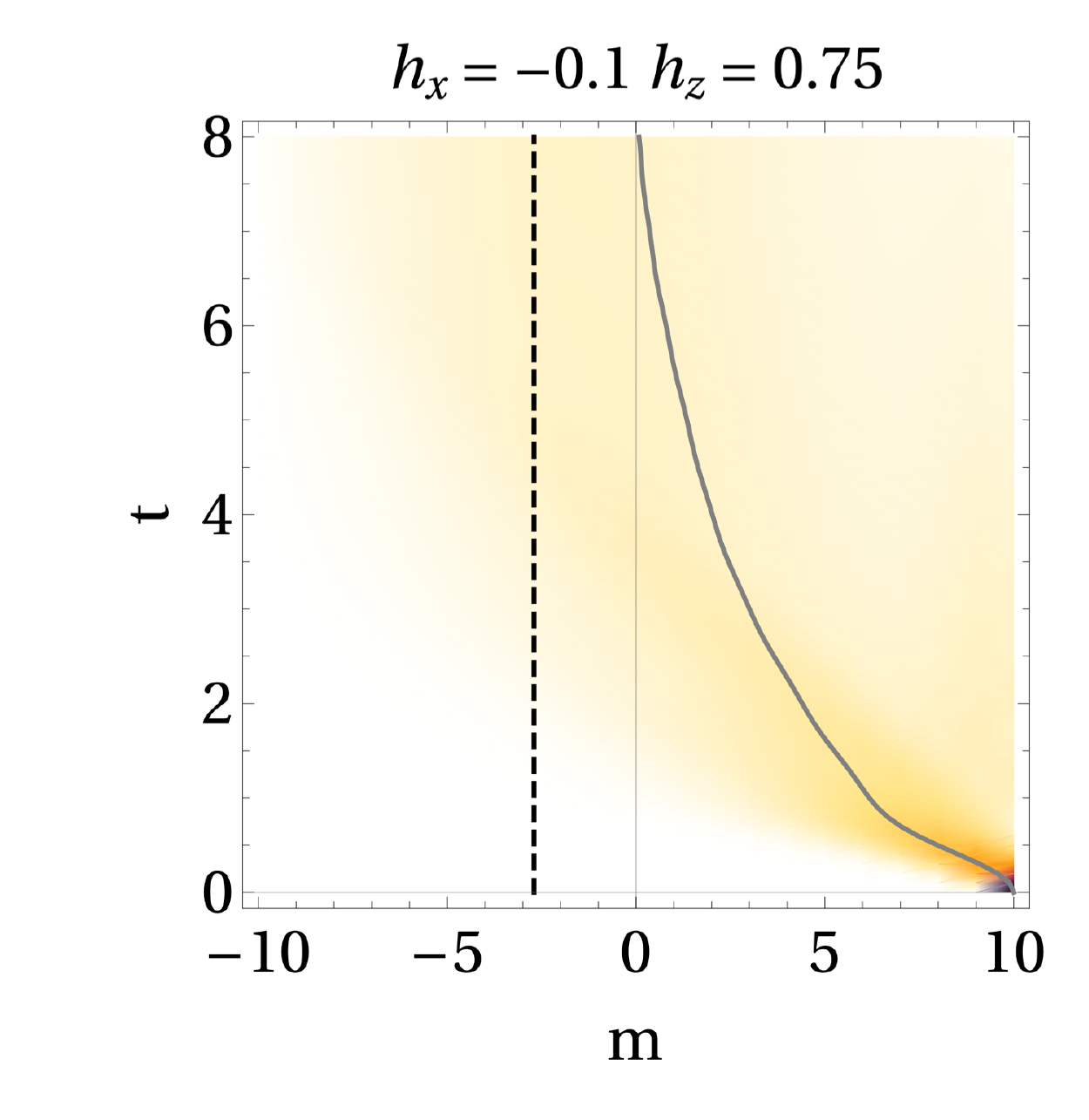}
\includegraphics[width=0.027\textwidth]{figs/fig4/bar.pdf}
\caption{
The same as in Fig.~\ref{fig:pdf_para}
for quenches to the  ferromagnetic phase and subsystem size $\ell = 20$.
}
\label{fig:pdf_ferro}
\end{center}
\end{figure*}

When quenching to the ferromagnetic region, the confinement strongly affects the relaxation 
~\cite{bib:kctc17,bib:jkr19,bib:rjk19,bib:mplcg19,bib:lzsg19,vk-20,bib:lsmpcg_arxiv}.
A representative set of our results for the dynamics of the order parameter statistics are in Fig. \ref{fig:pdf_ferro} for $h_x =\pm 0.1$ and $h_z = 0.5$ and $h_z=0.75$.
These parameters have been chosen to {\it not} be deeply in the ferromagnetic/confining phase, but still we observe many non-thermal features. 
While the precise value of $h_z$ only quantitatively changes the dynamics, the sign of $h_x$ provides very different features which are very nicely captured by the PDF.
This difference stems from the fact that for $h_x>0$ the initial state has energy close to the ground state, while for $h_x<0$ it is in the middle of the many-body spectrum
(and naively and erroneously one could expect to thermalise quicker). 

For $h_x>0$, the PDF remains localised around the initial value $\ell/2$ which is not far from the asymptotic mean value.
Anyhow, neither the shift toward the mean value nor the broadening is uniform  and the PDF has more than one maximum sometimes. 
The magnetisation shows the known oscillations related to the masses of the characteristic bound states~\cite{bib:kctc17,d-14}. 
The same oscillations are present in all cumulants, explaining the non-uniform relaxation.  

%When the initial polarisation is aligned with the confining field, the thermalisation dynamics is  almost ``frozen'':
%The typical relaxation time is expected to scale exponentially with the inverse of the quasi-particle density~\cite{bib:lsmpcg_arxiv};
%however, it is difficult to establish whether the slowing-down is mainly due to the large overlap with the ground state of the post-quench Hamiltonian. % (see previous paragraph). 

However, the strongest qualitative evidence for the lack of thermalisation appears when quenching to $h_x<0$.
In this case, the initial magnetisation is $\ell/2 > 0$, while the thermal distribution is centred around a negative value.
For small $h_z$, there are few dilute and heavy excitations (mesons); they cannot scatter and so the PDF is stuck 
on the initial {\it wrong side} of $m$, likely for an exponentially long time \cite{bib:lsmpcg_arxiv}.
This is visible in Fig.~\ref{fig:pdf_ferro} for $h_x =- 0.1$ and $h_z = 0.5$.
When the transverse field is increased, mesons become lighter (so larger typical size), their density larger and they may be produced 
in pairs of opposite momenta. Consequently they can scatter and drift toward equilibrium. 
(In other words, some mesons have size larger than the localisation length $\xi_{loc}\sim h_z /|h_x|$  \cite{bib:lsmpcg_arxiv}.)
%this density of excitations spoils partially the quasi-localised dynamics and 
Indeed a slow drift of the PDF toward the {\it right side} of $m$ may be seen  
%thermal equilibrium appears 
in Fig.~\ref{fig:pdf_ferro} for $h_z = 0.75$ and $h_x =- 0.1$ (still, at the larger accessible time, it is very far from thermal PDF).
However, also at these values of $h_z$ there exist heavier dilute mesons that can have a considerable overlap with the initial state;
they will remain localised for an exponential time. 
In conclusion, the interplay between the density of mesons and their masses (i.e. between the strength of the quench
and the confinement) manifests in a two-step relaxation of the PDF of  $M_\ell$:
{\it (i)} the lightest excitations may be abundantly produced (even in pairs) and cause part of the PDF to relax with a time $\propto\ell$;
{\it (ii)} the dilute heaviest excitations can only melt after an exponential time causing the freezing of another part of the PDF. %,  which therefore remains frozen for very long times. 

The phenomenology we just described for $h_x<0$ is just a dynamical manifestation of the Schwinger effect \cite{s-51}, i.e. the decay of the false vacuum, 
as already described for other observables \cite{bib:lsmpcg_arxiv,vlwgh-20}.
Such decay takes place via tunnelling effect for the production of pairs of mesons which is exponentially suppressed exactly as outlined above.

 \section{Conclusions}
We studied the relaxation dynamics of the order parameter statistics in the quantum Ising chain under magnetic confinement.
We inspected different regimes, by varying the Hamiltonian couplings.
 We found that a finite value of a longitudinal field may affect 
the gaussification of the order-parameter distribution function, 
depending whether it is tuned in the same direction of the initial polarisation or not.
This phenomenon is somehow related to the memory of the original local order.
 
We showed that the PDF is an ideal quantity to qualitatively show the lack of thermalisation (in numerically accessible time windows)  
in the presence of confinement, despite the non-integrability of the model.
In particular, we argued that an eventual relaxation can happen in a two-step process with the lightest mesons relaxing quickly and the
heaviest ones remaining frozen (likely for an exponentially long time). 
It is natural to wonder whether similar PDFs could shed some light also for other models that tend to avoid thermalisation, 
like those with quantum scars \cite{scar}.

\acknowledgments 
We thank Alessio Lerose for insightful comments. 
PC and RJVT acknowledge support from ERC under Consolidator grant  number 771536 (NEMO).

\end{document}